# NEREID: LIGHT CURVE FOR 1999-2006 AND A SCENARIO FOR ITS VARIATIONS

**Bradley E. Schaefer** (Department of Physics and Astronomy, Louisiana State University, Baton Rouge Louisiana 70803 ),
**Suzanne W. Tourtellotte** (Department of Astronomy, Yale University, New Haven Connecticut 06511),
**David L. Rabinowitz** (Department of Physics, Yale University, New Haven Connecticut 06511),
**Martha W. Schaefer** (Department of Geology and Geophysics, Louisiana State University, Baton Rouge Louisiana 70803 )

## ABSTRACT

Nereid is a small irregular moon of Neptune that displays large-, moderate-, and small-amplitude photometric variations on both fast and slow time scales. The central mystery of Nereid is now to explain the physical mechanism of these unique brightness changes and why they change with time. To characterize Nereid's variability, we have been using the SMARTS telescopes on Cerro Tololo for synoptic monitoring from 1999 to 2006. We present a well-sampled photometric time series of 493 magnitudes on 246 nights mostly in the V-band. In combination with our earlier data (for 774 magnitudes over 362 nights), our 20-year data set is the most comprehensive for any small icy body in our Solar System. Our yearly light curves show that Nereid displays various types of behaviors: large amplitude brightenings and fadings (1987 to 1990); moderate-amplitude variation about the average phase curve (1993-1997, 2003, 2005), moderate-amplitude variation and systematically brighter by roughly one-quarter magnitude throughout the entire season (2004); and nearly constant light curves superimposed on a surprisingly large-amplitude opposition surge (1998, 1999, 2000, 2006). Other than in 2004, Nereid's variations were closely centered around a constant phase curve that is well fit with a Hapke model for the coherent backscattering opposition surge mechanism with angular scale of 0.7°±0.1°. In our entire data set from 1987-2006, we find no significant periodicity. We propose that the year-to-year changes in the variability of Nereid are caused by forced precession (caused by tidal forces from Neptune) on the spin axis of a nonspherical Nereid, such that cross-sectional areas and average albedos change as viewed from Earth.

## 1 Introduction

Nereid is a small moon with an unusual orbit around Neptune (e=0.75, P=360 days, and i=28°). Nereid was viewed from afar (4,700,000 km) by the Voyager 2 spacecraft, so the best image is only 9x4 pixels and has been interpreted to be a sphere with average radius 170±25 km (Thomas, Veverka, and Dermott 1980). Nereid might be a captured Kuiper Belt object (Schaefer & Schaefer 1995; 2001) or it might be an original inner moon of Neptune nearly ejected by Triton (Goldreich et al. 1989).

Nereid has the most unusual photometric history of all objects in the Solar System. Photometric variations were discovered in 1987 (Schaefer & Schaefer 1988) with large brightness changes detected in multiple bands on the time scale of hours. Such large changes were later confirmed by several groups over many observing runs with data up through 1990 (Bus, Larson, & Singer 1988; Bus & Larson 1989; Williams, Jones, & Taylor 1991; Schaefer & Schaefer 2000). From 1993 to 1997, Nereid's amplitude was smaller by ~0.4 mag (Schaefer & Schaefer 2000). Two groups (who were each able to observe Nereid on just two consecutive nights) reported 0.14 mag and 0.09 mag variability over 24 hours at the 4.7-sigma and 5.4-sigma confidence levels respectively (Buratti, Goguen, and Mosher, 1997; Brown and Webster 1998). In 1998, we observed Nereid on 52 nights and found no significant variability other than a very large opposition surge (Schaefer & Tourtellotte 2001).

Recently, Grav, Holman, & Kavelaars (2003) have reported a photometric modulation with a period of 0.480±0.006 days (or half that period) and an amplitude of 0.029±0.003 mag. This is based on three half-nights of data plus short stretches of data on two other nights (in August 2001 and August 2002). We take this period to be 'unproven'. One problem that worries us is that we do not see how their observations can have one-sigma error bars of typically 0.003-0.006 mag for such a faint object. Systematic errors in CCD chips can not be made smaller than ±0.01 mag without heroic measures, and Buratti, Goguen, and Mosher (1997) used longer exposures on a larger telescope yet were only able to get one-sigma error bars of roughly ±0.04 mag. Another worry is that the claim for such high accuracy then suddenly makes small effects important, and the paper did not relieve fears that such effects were accounted for to high precision. A specific example is that their stated analysis procedure (Howell 1989) does not allow for the 8% changes in the photometric zero-point across the field arising from variable pixel scale (Schommer et al. 2000) and how this would change their differential photometry with respect to comparison stars whose positions on the chip vary through the night and night-to-night. Because we do not have adequate answers to these technical worries, we remain cautious about acceptance of the claimed periodicity. Although the claimed periodicity is plausible, a confirmation is needed.

How out-of-round is Nereid? This will become a central question for models to explain Nereid’s variability. Only two types of information are currently available to address this question; the single poor picture from Voyager 2 (see Fig. 1) with Nereid appearing ~9x4 pixels in size and a general comparison with the shapes of other moons of similar size. Both of these sources of information allow for Nereid to have axial ratios from roughly 1.08:1 (i.e., ~8% out-of-round) to 2.4:1 (see Section 7).

Because of Nereid's long history of changes in its variability, it is prudent to continue photometric monitoring. Ideally, we would like frequent magnitude measures throughout the year over many years. Such needs are completely outside the realm of possibility for most telescopes where a program might get a few nights every season. Fortunately, we have a long running program (Rabinowitz, Schaefer, and Tourtellotte 2007) for synoptic observing of small icy bodies in the outer Solar System with the SMARTS telescopes (Bailyn 2004) at the Cerro Tololo Inter-American Observatory (CTIO) in Chile. The four SMARTS telescopes (the CTIO 0.9, 1.0, 1.3, and 1.5 meter telescopes) are operated in a queue-mode, wherein a resident observer takes observations for many programs every clear night. This allows individual programs (like ours) to nightly monitor a target all year and year-after-year. As such, we have a unique program that produces long-term light curves perfectly suited to many aspects of Nereid investigations.

## 2 Photometry

As part of our continuing program of Nereid photometry, we have been monitoring the moon on many nights from 1999 to 2006. A journal of our observations is presented in Table 1. All of our observations were 15-minute exposures made with the SMARTS telescopes at CTIO. We have good coverage during 6 years in the V-band. In 2003, we also have good coverage in the B-band and I-band. We have also included 23 unpublished R-band magnitudes (from 13 nights) observed by A. Fitzsimmons and R. T. J. McAteer in August 1999 (JD 2451394-2451406) with the 1.0 meter Jacobus Kapteyn Telescope on La Palma.

Our data processing and reduction techniques are presented in detail in Rabinowitz, Schaefer, and Tourtellotte (2007), Schaefer & Tourtellotte (2001), and Schaefer & Schaefer (2000). This includes the examination of each image for cosmic rays, image defects, and nearby bright stars, as well as the examination of other images of the same star field after Nereid has passed along for the presence of any star or galaxy within the photometry aperture. Nereid moves about 1 arc-second in our 15-minute exposure intervals, and this is sufficiently smaller than the diameter of our photometry aperture (typically 3.5”) and the seeing disk size (typically 1.5”-2”) such that the effects of Nereid’s motion are negligible. We have found from broad experience with many stellar and Solar System targets that our photometry has a systematic uncertainty which appears to be constant at the 0.015 mag level, and so we have added this in quadrature with the statistical uncertainties so as to get the final total error bars. (These statistical uncertainties are calculated in the standard way with the *Image Reduction and Analysis Facility*, *IRAF*, software package.) Some of our data images have valid data for which the one-sigma uncertainty is larger than 0.2 mag (for example, due to thin clouds or a bright nearby Moon), and these 47 points are not included in our analysis. In all, the median one-sigma error bar for our 383 V-band observations is 0.05 mag.

Our individual magnitudes are presented in Table 2. The first column is the Julian Date for the middle of our exposure for the time when the light left Nereid. That is, we report $JD_{mid}-\Delta*0.005776$ where $JD_{mid}$ is the Julian date of the middle of our exposure and '$\Delta$' is the Earth-Nereid distance in Astronomical Units. The second column is the band for the magnitudes, with most observations being in the V-band. The third column is the solar phase angle $\phi$ for Nereid, that is, the Sun-Nereid-Earth angle in degrees. The fourth column lists the standardized magnitudes for Nereid and their one-sigma error bars. The standardization is to correct our observed magnitude ('m') for the varying Sun-Nereid distance ('r' in AU) and the varying Nereid-Earth distance ('$\Delta$' in AU), where we tabulate m-5*$Log_{10}$[rΔ/900]. Our choice of using the constant '900' is so that the tabulated magnitudes are fairly close to those actually observed. Absolute magnitudes for Solar System bodies take the convention of r=Δ=1AU, which gives a constant offset from our standardized magnitudes of 5*$Log_{10}$[900]=14.77 mag brighter. Thus, at zero phase with a standard magnitude of 19.18 mag, Nereid will have an absolute magnitude of 4.41 mag. We have taken the values of $\phi$, r, and Δ from the HORIZONS ephemeris program (http://ssd.jpl.nasa.gov/horizons.cgi) provided by the Solar System Dynamics group at the Jet Propulsion Laboratory.

In all, we are reporting on 493 observations on 246 nights over six years between 1999 and 2006. This is to be added to our previous work that reports 224 magnitudes on 64 nights from 1987 to 1997 (Schaefer & Schaefer 2000) and 57 V-band magnitudes on 52 nights in 1998 (Schaefer & Tourtellotte 2001). The BVI observations from 2003 have already been discussed in Rabinowitz, Schaefer, and Tourtellotte (2007). In all, our long-term monitoring program of Nereid now reports on a total of 774 magnitudes on 362 nights over 20 years.

**3 Fitting the Phase Curves**

A good way to organize our Nereid photometry is with its phase curve for a given apparition. Thus, we present yearly plots of the standardized V-band magnitude versus solar phase angle in Fig. 2. For comparison, we have also included the phase curves from 1987 to 1990, 1993 to 1997, and 1998 as taken from our previous work (Schaefer & Schaefer 2000, Schaefer & Tourtellotte 2001). The plots are all on identical scales to allow easy comparison.

The plots all show systematic variations as a function of the phase angle. In particular, many of the years show phase functions that are steeper at small phase angles than at large phase angles. Schaefer & Tourtellotte (2001) fitted the 1998 data and proved that the slope at low phase angle was steeper than at high phase angle. A perusal of Fig. 2 shows that the phase curve appears to be roughly constant from year-to-year, and the possibility of a constant phase curve is alluring for its simplicity.

To quantify this constant phase function, we have selected out the years with minimal variability and created a combined phase curve (see Fig. 3). We selected years (1998, 1999, 2000, and 2006) with good phase coverage and RMS variations ≤0.06 mag (cf. Table 4). For these selected years, the magnitudes from each year are lying on top of each other to within the given error bars. To quantify this phase curve, we have sorted by phase the measured V-band magnitudes for all four years and performed a weighted average (with associated error bar) over all points within narrow phase ranges. This average phase curve is presented in Table 3, with the first column being the average solar phase angle, the second column being the range of phase used in the weighted average, the third column being the number of magnitudes in the average, and the fifth column being the average distance corrected magnitude. The last column gives the slope of the phase curve between the two adjacent lines. With this, we see that the underlying phase curve has continuously changing curvature.

We are wanting to fit our observed phase curves to a physical model, both because we can derive the associated physical parameters for Nereid, but also because a physical model will provide the best functional form for characterizing our data. Fortunately, Hapke (1993; 2002) has presented a very detailed model that includes a wide range of physical effects as for how the brightness of an object will change as a function of all the angles involved. In essence, Hapke models the full radiative transfer of sunlight falling on a surface, including the effects of surface roughness, the phase function for individual scattering sites, macroscopic and microscopic shadowing, and interference effects. What is relevant for this paper is that Hapke presents specific models for the two physical mechanisms that cause opposition surges; shadow-hiding and coherent backscattering. Shadow-hiding is the effect that a surface at zero phase angle will appear brighter than at higher phase angles due to the lack of shadows because the shadows are all hidden behind whatever is casting the shadows. Coherent backscattering is the effect where a surface at zero phase angle will appear brighter than at higher phase angles due to constructive interference between light waves following identical paths in reverse from closely spaced scattering sites.

Hapke (2002, equation 38) presents a specific equation for the opposition surge that includes the effects of both shadow-hiding and coherent backscattering. In fitting the Nereid phase curve from Table 3, the relevant parameters are the amplitudes and angular widths of both surge mechanisms. (Essentially identical fits are found by using the full set of unbinned data from the four years.) For the coherent backscattering mechanism alone (i.e., with $B_{S0}$=0 in Hapke's terminology), we find a good fit when the angular width (i.e., $h_C$ in Hapke's terminology) is 0.77°±0.07° (0.0134±0.0012 rad) and the amplitude (i.e., $B_{C0}$ in Hapke's terminology) is 0.54±0.02. With the measures of the phase curve from the previous paragraph, we get a chi-square of 7.5, which indicates that Hapke model fits our data to typically 0.016 mag. This best fit Hapke model has an equation for the standard magnitude as

$$m_{opp} = 19.655 - 2.5*\mathrm{Log}_{10}\{1 + 0.54\ [1+(1-e^{-z})/z]/[2(1+z)^2]\}, \qquad (1a)$$

$$z = \mathrm{Tan}[\phi/2]/(0.0134) \qquad (1b)$$

for a solar phase angle of $\phi$. The quantity $m_{residual}$=m-5*$\mathrm{Log}_{10}[r\Delta/900]$-$m_{opp}$ will be the observed magnitude corrected for all effects due to distances and geometry.

For the shadow-hiding mechanism alone, we find no satisfactory fit even if we push the width to non-physical small values. For $B_{C0}$=0, the best fit is with $B_{S0}$=1 and $h_S$=0.2° for a chi-square of 41.8.

We find that a range of solutions is possible with nearly the same coherent backscattering contribution for any amount of contribution by shadow-hiding. For example, we also get a good fit (with chi-square equal 5.7) for a maximal shadow-hiding contribution (with $B_{S0}$=1) of angular width $h_S$=7° plus a coherent backscattering having a width of $h_C$=0.65°±0.07° (with $B_{C0}$=0.45±0.02). What is going on is that the shadow-hiding contribution is always too broad to change the steep rise at low angles, so the only way to account for our observed steep rise is to attribute it to coherent backscattering. That is, we cannot constrain the amplitude of the shadow-hiding contribution, but the steep rise to low phases is certainly caused by coherent backscattering at about 50% of its maximum possible value. Thus, we take our 'constant' Nereid phase curve as strong evidence for the existence and dominance of coherent backscattering.

The best fit parameters for the coherent backscattering depend somewhat on the amount of shadow hiding, which we cannot constrain. We can represent the situation by quoting the middle values between the extreme cases along with an error bar that covers these case, so we give $B_{C0}$=0.50±0.05 and $h_C$=0.7°±0.1°.

We recognize two anomalies between our data and *early* predictions based on coherent backscattering models (Hapke 1993; Mishchenko 1992). First, the albedo of Nereid is low (0.19, Thomas, Veverka, and Helfenstein 1991) so only a small fraction of the light is multiply scattered and hence coherent backscattering cannot be significant. But this anomaly is resolved by the discovery that low-albedo objects have strong coherent backscattering (Hapke et al. 1993; 1998; Helfenfenstein, Veverka, and Hillier 1997) and theoretical consideration of scattering *within* a single particle shows that coherent backscattering can be strong for light that is effectively singly scattered (Hapke 2002). Second, the width of the phase function does not vary linearly (or even substantially) with wavelength as predicted (Hapke 1993). In particular, the slopes at low phase angles for B-, V-, and I-bands are the same, with the differences in mean magnitude for phase ranges 0.00°-0.15° and 0.28°-0.54° being 0.14±0.01 mag for all three bands. But this anomaly is resolved with experimental measurements that show the width of the opposition surge to have only a weak dependence on the ratio of the wavelength to particle size (Nelson et al. 2000).

We can take a smooth continuously-curving Hapke model (equation 1) as a constant that should be applicable to all years. The residuals from this model will then be the intrinsic variability of Nereid. Our Hapke model from equation 1 (with no shadow-hiding, $B_{C0}$=0.54 and $h_C$=0.77°) is overplotted in all the panels of Fig. 2. We also plot the residual magnitudes ($m_{residual}$, corrected for all effects of distances and geometry) for all data from 1987-2006 in Figure 4. We can also calculate the chi-square values for a comparison between our observed magnitudes and the Hapke model. The results of these comparisons are given in Table 4. The first column is the year(s) of the fit, the second column is the chi-square of the fit along with the number of degrees of freedom, the third column is the RMS scatter of the residuals between the model and data, and the fourth column is the average residual (with a negative sign indicating that Nereid appears brighter than the model). The final column is a verbal description of the variations in the observed magnitudes when compared to the Hapke model.

The Hapke model for the coherent backscattering opposition surge on Nereid provides a good fit to the data. We take this as a strong indication that Nereid's surge is dominated by coherent backscattering. The relatively low albedo for Nereid and the lack of color dependency in the surge width provide non-laboratory evidence that the two *early* predictions for coherent backscattering are not required in real bodies. The constancy at which Nereid follows our single best fit Hapke model for many years suggests that the basic phase curve (and hence the microscopic texture on the surface) remains constant.

## 4 Does the Phase Curve Change?

The phase curve is produced by the microscopic 'texture' of the surface of Nereid. This could well be independent of whatever mechanism is giving rise to the variations about that phase curve. If so, then the underlying phase curve should remain constant. This would be apparent by Nereid always returning to the same phase curve whenever variability is low. Indeed, Fig. 2 displays a fairly convincing case for just this situation and conclusion. Numerically, we can see this in Table 4 because all the years (except 2004) have a near-zero average offset from the Hapke model. Further, the years with little variability have small RMS variations and moderate chi-squares. In all, it appears that the Hapke model phase curve is the center of variations which is constant throughout two decades.

But we are not so confident of this conclusion when we look in detail at the chi-squares for the fits. In particular, fits to phase functions with adjustable parameters (for which we have adopted a simple broken line) lead to smaller reduced chi-square values. In most of the years, the chi-square from the best-fit broken-line is numerically significantly smaller than that obtained with the Hapke model. With this, apparently, the year-to-year phase function varies somewhat around the Hapke model. For this difference, we have to account for the Hapke model (i.e., Eq. 1) being fit to each year's data with no adjustable parameters whereas the broken-line model has four adjustable parameters. So the question is then whether the use of four additional adjustable fit parameters (with their inevitable improvement in the chi-square) can account for the improvement? This can be answered with an F-test (Bevington and Robinson 2003, page 207). For four additional fit parameters, the F statistic will be the $F=[(\chi^2_{Hapke}-\chi^2_{BL})/4]/[\chi^2_{BL}/(N_{obs}-4)]$, where $\chi^2_{Hapke}$ is the chi-square for the Hapke model, $\chi^2_{BL}$ is the chi-square for the broken-line model, and $N_{obs}$ is the number of observations (which equals the number of degrees of freedom as given in Table 4). The probability distribution of F will be $P_F(F;4,N_{obs}-4)$ and can be looked up in standard tables. For $49<N_{obs}<94$, the 5% probability level has $F\approx2.5$, while the 1% probability level has $F\approx3.7$. We find that the years 1987-1990, 2000, 2003, and 2005 all have small F, which implies that the smaller chi-square of the broken model is understandable simply because it has more adjustable fit parameters than does the Hapke model. The years 1998, 1999, and 2006 have F values of 7.2, 6.5, and 5.6 respectively, with the implication that the non-Hapke model is preferred (and not just simply because it has more adjustable parameters). So for these three years, an F-test case can be made that Nereid's phase curve was *not* constant.

Does Nereid have a constant phase curve? We have just presented a plausible case that it is indeed a constant, and also a plausible case that there are small variations in at least a few years. How can we reconcile these two arguments? We think that the reconciliation lies in the fact that Nereid does have large and small amplitude variations superimposed on the phase curve and that these variations can easily be interpreted as changes in the underlying phase curve. That is, for example, in 1998 Nereid suffered some small changes in brightness (by some mechanism other than a changing phase curve) that produced an observed phase curve with small differences from the constant phase curve leading to a relatively large F value and the suspicion that the phase curve is variable. Such an argument is 'slippery', because it can be applied in all situations and even true cases of changing phase curves can then be covered up as arising from some separate mechanism. However, given that Nereid does indeed have large, moderate, and small amplitude variations on time scales faster and slower than any possible variations due to a phase curve change, we find it plausible that the slight variations about the Hapke model are entirely caused by unrelated variations. As such, we conclude that Nereid apparently has a constant phase curve, and that this phase curve is dominated by coherent backscattering.

## 5 Search for Periodic Variations

Given Nereid's small size, it must be inevitably out of round at some level. In addition, albedo variations across its surface are likely and were seen with Voyager 2. As Nereid must rotate, we expect to see periodic rotational modulation of its brightness.

We can use our extensive photometry to seek periodic brightness changes. Indeed, our unique many-year nightly photometry (with 774 observations in all bands on 362 nights) is well-suited to the search for the long periods as required by chaotic rotation. And the large number of our observations allows us to look for even short periodicities that are coherent. To this end, we have used our Hapke phase curve to create a set of residuals from which all the geometrical effects have been removed.

The residual magnitudes have been subjected to a discrete Fourier transform program which we have written in *Mathematica*. The probability of random residuals returning a Fourier peak with power P (in units of the average power in the transform) or more anywhere in the transform is $N_{freq}*e^{-P}$, where $N_{freq}$ is the number of independent frequencies in the transform. $N_{freq}$ is the number of frequencies throughout the examined range that are uncorrelated, with this being a problematic quantity for nonuniformly sampled data. Approximately, $N_{freq}=(\nu_{max}-\nu_{min})(T_{end}-T_{start})/(2\pi)$, where the data ranges in time from $T_{start}$ to $T_{end}$ and the Fourier Transform ranges between frequencies $\nu_{min}$ to $\nu_{max}$ (Frescura, Engelbrecht, and Frank 2007) For the highest peak in a transform to be considered significant, we require that this probability be much less than unity and in particular that the probability be less than the 3-sigma value of 0.0027. We have searched for periods from 0.1 days to 50 days. No peak in the Fourier power spectrum stands out above the usual noise peaks and no peak approaches the 3-sigma level. That is, we find no significant periodicities.

We can use our data to place constraints on the possible amplitude of a periodic modulation. This was done by computing the sine wave amplitude that returns a Fourier power of $Ln(N_{freq}/0.0027)$ in units of the average power. This can be done either by direct computation from the Fourier Transform with random phases or by superposing a sine wave of known amplitude on the data itself. For our typical one-year observing season, the period range of 0.1-50 days results in ~1400 independent frequencies. Our 3-sigma limits on the peak-to-peak amplitude for each year of our observations are presented in Table 5. These limits are close to twice the RMS scatter (see Table 4) and this is not coincidental because of our sparse sampling.

We have also made Fourier transforms of other intervals, such as 1998-2000, as well as the entire data set from 1987-2006, and we have looked for sub-threshold peaks that appear at the same frequency in multiple years. As a rigorous means of looking for periodicities that are incoherent from year-to-year (i.e, with the period or phasing changing slightly between years), we have used the technique of coadding the Fourier power spectra (Schaefer et al. 1992) so that real peaks will add linearly but noise peaks will not rise fast as more years are combined. Again, we find no significant periodicities.

An important specific test is to look for the claimed periodicity of Grav, Holman, and Kavelaars (2003). We do not have any observations in the years 2001 and 2002, but their periodicity would be expected to be coherent and present in other years. They quote their period as 0.480±0.006 days for an *assumed* double-peaked light curve, even though they show no difference between odd and even numbered peaks. For a Fourier transform period search, we must look at the simple sine wave period of 0.240±0.003 days. The number of independent frequencies over this range is 100 times smaller than $N_{freq}$ for the full 0.1-50 day range, so the limiting amplitudes can be pushed to lower levels. Our 3-sigma upper limits are presented in Table 5. Our lowest 3-sigma limits are for a peak-to-peak amplitude of 0.07 mag. That is, our data are not sensitive enough to detect oscillations with a full amplitude of 0.029 mag such as claimed by Grav, Holman, and Kavelaars (2003). Nevertheless, we have looked around the claimed period in our Fourier transforms for various time intervals. From this examination, we find no evidence of the claimed periodicity (as expected due to our sensitivity).

Nereid must have *some* rotation period (even if the rotation vector changes in magnitude and direction over time) and for an out-of-round moon (see Section 7) there must be an inevitable photometric modulation. *If* our observed brightness variations are associated with the rotation of Nereid, then we can place two useful constraints on the rotation period even though we do not have a formal periodicity. First, we would find a contradiction with the claimed 0.480 day (or 0.240 day) rotation, because Nereid cannot suffer both very-small-amplitude rotational modulation at the same time as a large-amplitude rotational modulation. Second, the rotation period must be shorter than roughly 3 days or so. The basis of this is that we see fast variations (intranight and between successive nights). For a sinusoidal shaped light curve with peak-to-peak amplitude of A, the spin period must be less than $\pi A/\mu$ (or $2\pi A/\mu$ for a double-peaked sinewave), where $\mu$ is the observed rate of change in the object's magnitude. The most restrictive period will come from the fastest variations. Over the night of 18 June 1987 (JD2,446,964), Nereid was seen to vary fast and consistently in all five UBVRI bands, with the average variations over the optical bands of $\mu$=0.20±0.05 mag hour$^{-1}$ (Schaefer and Schaefer 1988). For the observed amplitude in that month of A=1.2 mag, we estimate that the rotation period should be less than ~0.8 days (or ~1.6 days). For the new magnitudes reported in this paper, our fastest internight variation is between JD2453185-2453186 with $\mu$=0.15±0.03 mag day$^{-1}$ and A=0.23±0.07 mag, for $P_{spin}$<4.8±1.7 days (or $P_{spin}$<9.6±3.4). Our fastest intranight variation is on JD2453590 with $\mu$=0.32±0.09 mag day$^{-1}$ and A=0.23±0.05 mag over the season, for $P_{spin}$<2.3±0.8 days (or $P_{spin}$<4.6±1.6 days). With substantial uncertainties (both from the measurements and from the shape of the light curve), we adopt a rough upper limit on the rotation period of Nereid to be less than ~3 days or so.

If Nereid's rotation has significant changes due to precession (see Section 8), then the amplitude and character of Nereid's brightness modulations will change from year-to-year. The precessional changes need not appear exactly periodic, but we can nevertheless put constraints on the approximate precession period. From Table 4 or Figure 4, we see that Nereid undergoes years of relative calm and years of relative activity, with these episodes lasting several years. Nereid has low amplitude variations (beyond that of the average phase curve) in the years 1998-2000 and 2006. Nereid is active in the years 1987-1990 and 2004-2005. We have little useful information for the years 1991-1997 and 2001-2002, while the light curve from 2003 has larger than desired uncertainties. If these episodes are roughly periodic, then we can place constraints on the period. The interval between the calm years (1998-2000 to 2006 and perhaps later) suggests a period of 6-8 years (with an unrecorded calm episode around 1991 or soon thereafter). The interval of active years (1987-1990 to ~2004) suggests a period of 14-17 years or ~8 years (with an unrecorded active episode around 1996). Taken together, the calm and active episodes give a period of ~8 years. On the possibility that the precessional modulation is double-peaked (due to episodes caused by first one hemisphere than the other pointing towards Earth), the precession period might actually be ~16 years. In all, the variation of Nereid's amplitude over the years suggests that its precessional period might be ~8 or ~16 years.

We have placed limits on periodicities in Nereid's light curve: (a) We find no significant periodicities from 0.1-50 days with limits in amplitude as low as 0.08 mag. (b) Our observations are not sensitive enough to test the small amplitude periodicity reported by Grav et al. (c) If our observed internight and intranight variations are caused by rotational modulation, the period of Nereid must be less than something like ~3 days. (d) If the year-to-year changes in the amplitude of modulations are caused by precession, then Nereid's precessional period is apparently ~8 or ~16 years.

## 6 Brightness Changes Superposed on the Phase Curve

Fig. 2 and Table 5 show that Nereid often has variability about the constant phase curve and that these variations are both large and small in amplitude and are changing over time. This is the central mystery of Nereid.

The first question to be answered is whether the variations are due to instrumental problems. Schaefer and Schaefer (2000) and Schaefer and Tourtellotte (2001) addressed this question in detail. The strong conclusion is that the variations are certainly not instrumental for any of four reasons. Briefly, the reasons are (1) five groups of researchers over eight observing runs see the large-amplitude variations, (2) significant hour-to-hour trends are seen in UBVRI data from one night, (3) the deviations are both positive and negative thus ruling out background stars and cosmic rays, and (4) the large-amplitude variations are greatly larger than is plausible for any error given the vast experience of photometry with the same telescope/filter/CCD/procedure combinations.

The light curve from 2004 displays a different type of variability, in that the entire light curve appears to be roughly a quarter-magnitude brighter than the constant phase curve. Could this offset arise from instrumental or calibration problems? Instrumental problems are very unlikely as we used two separate telescopes with the same result and concurrent programs on both showed no similar problems. For calibrations, we independently measured the comparison stars in several-day groups with all groups being directly calibrated against standard stars (Landolt 1992) on many photometric nights. As such, a systematic offset for an entire year would require not a single point failure but would require many-point failures of independent calibrations all in the same direction and offset. And as for internight variability, we have checked by looking at only differential photometry from identical groups of comparison stars. So for example, JD 2453185 (with 3 images) and JD 2453186 (with 6 images) share three comparison stars (which are stable relative to each to better than 0.02 mag) for which Nereid has changed by 0.15±0.03 mag. (Such variations over 24 hours have been confidently reported by many groups earlier, including Buratti, Goguen, and Mosher, 1997; Brown and Webster 1998.) Such night-to-night variations are of high confidence and impossible to explain by any plausible instrumental or analysis problems.

Grav, Holman, and Kavelaars (2003) have put forward the possibility that the large-amplitude variations seen by many groups from 1987 to 1990 were caused by superposition of Nereid's image on top of chance background stars. From 1987-1990 the galactic latitude of Nereid was -7° to -13°, whereas from 1993-2006 the galactic latitude was -17° to -32°. This idea of observational error is alluring because then no explanation is needed for the large amplitude variations. However, this idea is certainly wrong for several strong reasons. (1) Half the variations are for Nereid being *fainter* than the average phase curve (see Fig. 2, first panel). With the use of many comparison stars calibrated from Landolt standards, the idea of contamination from a background star *cannot* account for such *dimmings*. (2) With CCD images, the detection of background stars is easy. (In the old days of photoelectric photometers, this was indeed a hard problem, but such an antiquated bias is not applicable here.) The reader can closely examine Figs 1 and 2 of Schaefer & Schaefer (2000) to see the complete absence of any contamination in a case where Nereid was brighter by 0.24±0.06 mag on the first image as compared to the second. We have closely examined the positions of Nereid on independent images when Nereid has moved, and we know that background stars cannot account for the brightenings. (3) The brightenings are correlated on a time scale longer than the duration of Nereid passing over a star. With our 1.6 arc-sec aperture radius, any background star will take roughly 0.03 days to pass from one edge to the other edge. For two examples, on the nights of JD 2447715 and 2447716, Nereid was bright by 0.4 and 0.25 mag consistently over 5 and 7 images stretching over time intervals of 0.199 and 0.343 days respectively. Such runs of Nereid being bright can be explained as a straight-row of 5-10 equal-magnitude background stars that were not noticed, but this is not plausible. The idea that the large amplitude variations of Nereid were caused by superposed background stars cannot be true.

In all, we have many strong reasons from many observers to know that the Nereid does indeed have brightness variations about its average phase curve that change in character and amplitude from year-to-year.

So what causes the brightness variations of Nereid? (That is, past those produced by changing Earth-Nereid distance, Sun-Nereid distance, the coherent backscattering opposition surge, and possibly the shadow-hiding opposition surge.) The possibility of a binary moon (Farinella et al. 1989) was later disproved with Voyager 2 imagery. The possibility of episodic outgassing from Nereid cannot explain the *dimming* in brightness and has the substantial problem that the cold temperatures do not allow for outgassing (Prialnik, Brosch, and Ianovici 1995). (The outgassing from the similar-sized Enceladus is likely powered by tidal heating derived from its orbital resonance with Dione, whereas Nereid has no such resonance.) The only remaining idea is that the variations are caused by rotation.

Rotation will cause changes in Nereid's brightness due to either differences in albedo across its surface or due to different cross-sectional areas from different directions. If the rotational axis changes in direction from year-to-year, then the amplitude of variations can change year-to-year. As we shall see in Section 8, the changes in the rotational axis require that Nereid be substantially nonspherical in shape. So, we will revisit the question of Nereid's shape.

## 7 Shape of Nereid

How round is Nereid? That is, is the shape of Nereid spherical with some small surface relief (say, like the Earth's Moon), or is it greatly-nonspherical (say, like Saturn's Hyperion), or is it somewhere between? This question will be critical for evaluating whether Nereid's rotational axis can be precessing as an explanation of its changing variability. To answer this question, we have only two types of evidence. The first is the single resolved *Voyager 2* image taken from afar. The second is the knowledge of the shapes of other moons of similar size.

Nereid happened to be near apoapse at the time of the *Voyager 2* encounter with Neptune, so the best image was taken from a distance of 0.031 AU (12.2 times the Earth-Moon distance). At this distance, the pixel size corresponded to 43 km and the diameter of 340±50 km (Thomas, Veverka, and Helfenstein 1991) corresponded to 7.9±0.6 pixels across. With this, we see that deviations from a sphere of roughly 12% are easily possible. In addition, the poor signal-to-noise in the image can allow for substantially larger deviations. This limit only applies to the shape along the limb and terminator for that one image. With the solar phase angle being 96°, the constraints from the terminator profile are weak as most shape changes will not translate into shifts in its apparent position. Large deviations from a spherical shape are possible anywhere away from the half-circle of the limb at the time. That is, the Nereid image shows only a quarter of the moon and greatly-out-of-round features can be easily hiding. In all, the *Voyager 2* image only places a constraint of <12% for only a one half-circle, so the image does not really tell us whether Nereid is spherical or greatly nonspherical.

The lack of constraints from the *Voyager 2* image can be seen with a mosaic of moon images degraded to similar resolution (see Fig. 1). All nine images in the mosaic are of moons in the Solar System, with solar phase angle of around 96°, with a resolution such that the diameter is roughly 8 pixels across, and with the Sun illumination coming in from the left side. Some of the moons illustrated are spherical, others have 10-20% relief, while others are far from round (with axial ratios of 1.6:1 to 1.9:1). Readers are asked to look at Fig. 1 and decide which moons are spherical and which are nonspherical. Only after this exercise should the reader look at the key provided in Table 6. We find that our colleagues only have a ~50% likelihood of making a correct determination of whether one of the moons is spherical or not. In this case, we realize that the *Voyager 2* image does not place useful limits on how out-of-round Nereid can be.

With the limits from *Voyager 2* being so poor, we can turn to comparison with similar sized moons. Dobrovolskis (1995) has demonstrated that there is a size threshold for which moons must be close to spherical (due to self gravity), and that this threshold is close to the size of Nereid. The shapes of the three moons just-larger than Nereid and just-smaller than Nereid are given in Table 7. Nereid is flanked in size by moons that are out-of-round with axial ratios 1.08:1 (i.e., 8% out of round) to 1.8:1. But Dobrovolskis' size threshold is not sharp, as demonstrated by Hyperion and Proteus being far from spherical yet having their longest radii being substantially larger than Nereid's. In agreement with Dobrovolskis, we conclude that a moon the size of Nereid could be fairly round (say, with 1.08:1 axial ratios like Mimas) or greatly-nonspherical (say, with axial ratios up to 2:1 like for Hyperion).

Despite the wide range of possibilities given the available evidence, we can place likely limits. All of the middle-sized moons are substantially out-of-round with axial ratios at the 1.08:1 level or higher. And the largest axial ratio for any moon that we know of is 2.4:1 (for Atlas around Saturn). Nereid is unlikely to be far outside these bounds.

In all, what do we know about Nereid's shape? Both the *Voyager 2* image and the shape-size relation for other satellites allow for Nereid to be somewhere between roughly-spherical and greatly-nonspherical. The only way to get an answer will be to send a new spacecraft mission to Neptune.

## 8 Forced and Free Precession

Nereid is certainly rotating, and the very likely out-of-round shape and albedo variations will then force a photometric modulation. Such brightness variations must be present at some level. Ideally, rotation could provide a mechanism to explain all the variations, with no need to invoke any second mechanism. Previously, only two extreme rotation cases have been considered, that of simple rotation and that of chaotic rotation.

The simple rotation idea would be where the spin direction and period remain constant. This is exemplified by the reported 0.480 (or 0.240) day periodicity of Grav, Holman, and Kavelaars (2003). With a fast rotation period, the magnitude of the spin angular momentum will be sufficiently large that ordinary tidal torques cannot significantly change the spin orientation on any useful timescale. As such, the photometric modulation must remain constant and coherent throughout all of our data, and hence this idea cannot account for the year-to-year variations. Additional modulation mechanisms must then be present, and this separate physical mechanism would dominate over simple rotation. As such, the idea of simple rotation explains nothing of the central mystery of Nereid's variability.

At the other extreme is the idea that Nereid is chaotically rotating, with its spin axis direction, pole position, and spin period changing aperiodically on short time scales (Dobrovolskis 1995). This idea is alluring due to the precedent of the chaotic rotation of Hyperion (Klavetter 1989a; 1989b), as well as due to the apparently 'chaotic' variations observed on Nereid. Dobrovolskis (1995) points out that chaotic rotation is inevitable for Nereid in its highly eccentric orbit if (a) it departs by more than ~1% from a sphere and (b) its spin period is longer than 2 weeks. The first condition is very likely to be correct. The second condition does not agree with observations. In particular, the many observed cases of intra-night variations and the many more cases of significant variability over a substantial fraction of the total amplitude within two consecutive nights demonstrate a timescale that is much too short to allow for chaos (see Section 5). These observations indicate a rotation period of less than 3 days or so. As such, the alluring idea of chaotic rotation does not appear to be the case for Nereid.

Fortunately, there are two intermediate cases of rotation. Both of these cases involve precession of the spin axis of Nereid. In both cases, the spin axis presents a changing angle with respect to Earth so that the rotational amplitude changes year-to-year. One case invokes forced precession, where the gravitational pull of Neptune produces torques on Nereid that changes the direction of the spin with a period of a few years. The second case invokes free precession, where the spin vector is not along one of the principal axes of the Nereid's irregular shape resulting in a precession motion.

The following subsections will discuss the physics of both cases:

### 8.1 Forced Precession

If Nereid is nonspherical, then the gravitational pull from Neptune will produce torques on Nereid's spin, and this will make the spin axis precess. This is the same mechanism that causes the precession of the equinoxes as discovered by Hipparchus. With the spin axis changing orientation as viewed from Earth, we will be seeing Nereid changing in cross section and possibly in albedo, with the result that the average magnitude and the photometric amplitude would change from year-to-year. For example, the 0.24 mag brightening in 2004 could be caused by a rotational pole (with either a high albedo around the pole or a higher cross section as appropriate for rotation around the shortest axis) pointing towards us for a while.

As Nereid is almost certainly nonspherical, the only question then becomes the length of the precession period. If the precession period is too long, then this mechanism cannot explain the year-to-year variations. As discussed in Section 5, the long-term period in the amplitude is not well constrained, but we can at least conclude that the precession period must be longer than roughly 6-8 years.

The angular velocity of forced precession equals the torque divided by the spin angular momentum (e.g., Kleppner and Kolenkow 1973). The period of precession is then

$$P_{prec} = 4\pi^2 \, I_{Ner} \, \tau^{-1} \, P_{spin}^{-1}, \quad (2)$$

where $I_{Ner}$ is the moment of inertia of Nereid, $\tau$ is the torque of Neptune onto Nereid, and $P_{spin}$ is the spin period of Nereid. If Nereid spins too fast, then $P_{prec}$ will be too large to explain the changing variability.

The moment of inertia and torque both depend on the mass distribution in Nereid. The instantaneous magnitude of the torque is

$$\tau = G \, M_{Neptune} \, M_{Ner} \, \varepsilon \, R_{Ner}^2 \sin(2\Theta) \, D^{-3}, \quad (3)$$

where G is Newton's gravitational constant, $M_{Neptune}$ is the mass of Neptune, $M_{Ner}$ is the mass of Nereid, $\varepsilon$ is the fractional excess of Nereid's equatorial radius compared to its polar radius, $\Theta$ is the angle between the Nereid-Neptune vector and the spin vector, and D is the Neptune-Nereid distance.

The torque will change with the position of Nereid in its orbit. In part, this dependence with position in the orbit arises because the tidal forces are inverse-cube with a distance that changes by a factor of 7. In addition, the torque will change as the angle between Nereid's rotation axis and the Nereid-Neptune vector, $\Theta$, changes. If the obliquity of Nereid (that is, the angle between the Nereid rotation vector and the normal to Nereid's orbit) is zero, then the symmetry of the situation demands that the torque must also be zero (and hence no precession). If the obliquity of Nereid is 90° (i.e., Nereid's axis is in the plane of its orbit), then the torque will be zero for the two places in its orbit where Nereid's poles point towards Neptune. The maximal torque will occur for an obliquity of 45° when Nereid's pole points closest to Neptune at the time of periapse. The average torque must be integrated over the entire orbit.

Fortunately, Dobrovolskis (1995, equation 23) reports on the orbit-integrated precession rate with a simple expression. He gives

$$\omega_{prec} = -0.75 \, (n^2/\omega_{spin}) \, [(2C-B-A)/C] \cos(\zeta) \, (1-e^2)^{-3/2}. \quad (4)$$

Here, $\omega_{prec}=2\pi/P_{prec}$ is the rate at which the spin axis circles about the orbit normal, $\omega_{spin}=2\pi/P_{spin}$ is the spin rate for Nereid's rotation, and $n=2\pi/P_{orbit}$ is the orbital mean motion of Nereid around Neptune. Equation 4 is valid for the case where $\omega_{spin} >> n$, which is exactly the case of interest here for Nereid. The moments of inertia for the three principal axes are A, B, and C, with the rotation being around the C axis. The angle between the spin axis and the orbit normal is $\zeta$. The orbital eccentricity of Nereid is $e=0.75$.

Let us take the case where the equatorial radii are equal (A=B). Then,

$$P_{prec} = 2/3\ (P_{orbit}^2/P_{spin})\ [C/(A-C)]\ (1-e^2)^{3/2} / \cos(\zeta). \quad (5)$$

For $P_{spin,day}=P_{spin}/(1\ \text{day})$ and $P_{orbit}=360$ days, we get

$$P_{prec} = 68.4\ \text{years} / \{P_{spin,day}\ [(A/C)-1]\ \cos(\zeta)\}. \quad (6)$$

We further know that $\cos(\zeta)\leq 1$, $P_{prec}<16$ years or so, and $P_{spin,day}<3$. Then, we must have $|(A/C)-1| > 1.4$. This can only occur for a prolate structure (i.e., with A>C) with A/C>2.4 or so.

For an ellipsoid shape and uniform density, Dobrovolskis (1995) relates the physical radii along the principal axes (a, b, c) with the moments of inertia along these same axes (A, B, C). For the case with A=B, we have

$$A/C = (a^2+c^2)/(2a^2). \quad (7)$$

The axial ratio is c:a. For A/C>2.4, we require that the axial ratio be greater than 1.9:1. That is, for the forced precession model to work with $P_{prec}$ less than 16 years, we require the polar radius of Nereid to be something like twice at least as long as the equatorial radius.

How plausible is it that Nereid's shape has an axial ratio of greater than 1.9:1? We only need look at Hyperion to see that such a shape is reasonable. For the six non-Nereid moons in Table 7, two of them have axial ratios of 1.8:1 or greater. Of the five moons of Neptune just smaller than Nereid, 40% have axial ratios of 1.9:1 or greater (Karkoschka 2003). And the moons with axial ratios of greater than 1.7:1 in Figure 1 all have image shapes indistinguishable from that of Nereid. As such, it is reasonable to expect that Nereid is sufficiently nonspherical to allow for the forced precession model.

The forced precession idea is that Nereid has a precession period of 16 years or so, with its pole of rotation changing orientation with respect to Earth on that time scale. This would account for the year-to-year changes in Nereid's amplitude, with Nereid's apparent cross section and albedo changing with the pole shifts. The intra-night and night-to-night variations are caused by the ordinary rotation of Nereid as high-and-low albedo regions and large-and-small cross sections are presented to Earth. The precession period must be 8 or 16 years to account for the timescales of changing amplitudes, and the rotational period must be shorter than a few days to account for the intra-night and night-to-night variations. For these constraints on the periods, the forced precession model only works if Nereid is substantially out-of-round, with axial ratios of 1.9:1 or greater. A prediction of this model is that Nereid will continue to cycle through years of high and low variability as viewed from Earth.

## 8.2 Free Precession

If Nereid is not rotating along one of its principal axes, then it will undergo free precession. The idea is that Nereid's pole might swing around in a wide cone, so that different terrains and cross sections are presented towards the Earth resulting in rotational modulation that can change from year-to-year.

Free precession does not involve any torques, such as those from Neptune. With no torques, the total angular momentum vector and the kinetic energy of rotation must both be constant. This is achieved by the spin vector precessing around the angular momentum vector with a constant angle.

The free precession period, $P_{prec}$, is tied to a reference frame defined by the principal moments of inertia of Nereid, with the current 'north pole' moving around the surface. As the north pole moves around, a distant observer will see systematic changes in the average brightness and amplitude of variations on the time scale of $P_{prec}$, and we can attribute the year-to-year variations on Nereid to this precession. In addition, Nereid will also be spinning about its (changing) north pole with some period, $P_{rot}$. In an inertial frame (like that of distant observers), the observed rotation modulation will have a period of $(P_{prec}^{-1}+P_{rot}^{-1})^{-1}$. We can attribute Nereid's fast variations to this spinning as the visible albedos and cross sections change. For the observed intranight and internight variations (with $P_{rot}$<3 day or so) and the apparent long precession (with $P_{prec}$>8 years or so), the apparent period for the observed rotational modulation will be very close to $P_{rot}$.

In planetary astronomy, the best known example of free precession is the Chandler wobble, where the Earth's north pole moves in small circles with a period of around 433 days. Large-amplitude free precession has been discovered for Toutatis (Spencer et al. 1995) and Halley's Comet (Samarasinha and A'Hearn 1991).

The details of the rotation and precession vary strongly with the shape and spin of the rotating body (Samarasinha and A'Hearn 1991). Within the idea of explaining Nereid's year-to-year variations as a free precession period, we look to cases where the precession period is much longer than the rotation period. This happens when the body is only slightly nonspherical. To take a simple and characteristic case, let us take Nereid to be an ellipsoid where two of its principal axes have identical moments of inertia (I) and the third axis has a slightly different moment of inertia ($I_{pole}$). In this case, the free precession period is given as

$$P_{prec} = P_{rot}\,[I/(I_{pole} - I)] / \cos(\alpha) \qquad (8)$$

(Fowles and Cassidy 2005, p. 387). Here, $P_{rot}$ is the rotation period and $\alpha$ is the angle between the third principal axis and the axis of rotation. For our model of Nereid, we require that the spin axis be far from the angular momentum vector so that precession will allow large changes from year-to-year, hence we will take $\alpha$~45° as representative. If $P_{rot}$~1 day and $P_{prec}$~8 years, then we require $I_{pole}/I$ to be around 1.0005. The most extreme plausible case is with $P_{rot}$~3 day, $P_{prec}$~6 years, and $\alpha$~60°, for which $I_{pole}/I$ <1.0027. Thus, for the free precession model, we would require Nereid have its principal moments of inertia equal to within a quarter of a percent. This is not plausible given Nereid's small size and the precedent of other moons of similar size (see Table 7).

Another difficulty of the free precession model is that non-principal axis rotation is damped on a relatively fast time scale. What happens is that the changing spin direction causes the flexing of the body to change and this causes strains that dissipate energy. The flexing continues until the spin vector is aligned with one of the principal axes at which time the free precession stops. The time scale for the alignment of the spin axis with the principal axis has been calculated by Burns and Safronov (1973) and Harris (1994). Harris gives that the damping time scale for the wobble is

$$T_{damp} \sim \mu Q / (\rho\, K_3^2\, R^2\, \omega_{spin}^3). \qquad (9)$$

Here, $\mu$ is the rigidity of Nereid, Q is the quality factor, $\rho$ is the bulk density of the body, $K_3^2$ is a numerical factor related to the shape of the body, R is the radius of Nereid, and $\omega_{spin}$ is the angular frequency of rotation. Peale and Lissauer (1989) estimate that $\mu=4\times10^{10}$ dyne cm$^{-2}$ for ice and $Q\leq100$, for $\mu Q$ of order $10^{12}$ in cgs units. For Nereid, $\rho$~1.5 gm cm$^{-3}$ and R~170 km. $K_3^2$ is ~0.01 for a nearly spherical body (Harris 1994). With the observed intra-night and night-to-night variations, we have $P_{rot}$ less than ~3 days or $\omega_{spin}$>0.000024 s$^{-1}$. With this, we have $T_{damp}$ is less than half a million years.

So for the free precession model for Nereid to work, it must have been excited into non-principal axis rotation within the last million years or so. A variety of different mechanisms are known to excite non-principal axis rotation in Solar System objects. The Earth's Chandler wobble is excited by movements of the atmosphere and ocean pushing against the continents. The free precession of Halley's Comet is excited by the gas jets pushing in directions not pointing back to its center of mass. The free precession of Toutatis has a damping time comparable to the age of the Solar System, so its rotation could either be primordial or from an ancient collision. None of these specific mechanisms would work for Nereid. The only remaining plausible mechanism is for some body to strike Nereid an off-center blow.

A large body striking Nereid an off-center hit could have excited non-principal axis rotation. The physics of such an impact are presented in Burns and Safranov (1973). For an inelastic collision with impact parameter equal to Nereid's radius, a relative velocity of near Neptune's orbital velocity, and for Nereid spinning with a period of less than 3 days, we calculate that the impactor diameter must be larger than 24 km so as to have the change in the spin angular momentum be comparable to its pre-collision value. The probability of a >24 km diameter body hitting Nereid a near-the-pole collision in the last million year is very small. And such a hit is unlikely to leave Nereid spherical to within 0.05%.

In all, our free precession model for Nereid is possible, but it has two tremendous problems that are unlikely to be overcome. First, to get a sufficiently long precession period, Nereid must be spherical in shape to within ~0.05%. Second, to excite the non-principal axis rotation, Nereid must have been hit a glancing blow by a body of >24 km in diameter within the last million years. Given these two extreme problems, we cannot accept the free precession model.

### 8.3 Discussion

The idea behind our models is that a slow precession of the rotation axis can account for the year-to-year changes in Nereid's photometric behavior. That is, as Nereid's pole swings around, Earth observers will view different cross sections and regions with different albedos. Years with low variability might be when the poles are pointing close to Earth, while the quarter-magnitude brightening during 2004 might arise from a high albedo region being visible. The cycle of variations suggest a precession period of ~8 years or possibly (depending on the geometry) ~16 years. The observed intra-night and night-to-night variations require that the underlying rotation period be less than 3 days or so. Within this idea, we have an explanation for the central mystery of Nereid without having to invoke additional mechanisms.

The forced precession model can account for $P_{prec}$<16 years and $P_{rot}$<3 days only if Nereid is significantly out-of-round. For an ellipsoidal shape, the length must be longer than roughly 1.9 times the diameter of the round side. Such a shape is reasonable, given that other similar sized moons (e.g. Hyperion) have similar axis ratios. A substantial *perceived* problem is that the single Voyager 2 image is reported to be consistent with a spherical body. However, close study of Fig. 1 demonstrates that the poor resolution of the Nereid image allows for substantial deviations from a sphere to be hidden. Indeed, unless some analyst can reliably distinguish the spherical moons from the greatly-out-of-round moons in Figure 1, the conclusion can only be that the Voyager 2 image places no useful constraints on the shape of Nereid. Given that we cannot see any way to distinguish the moon shapes from Figure 1, we conclude that Nereid easily can be out-of-round at the 2:1 level and the forced precession model is reasonable.

The free precession model can work in theory, that is, we can get $P_{prec}$~8 years for $P_{rot}$<3 days. The two big problems are that the shape of Nereid would have to be nearly a perfect sphere (to within ~0.05%) for the precession period to be long enough and Nereid would have to have suffered an oblique impact by a body >24 km in diameter within the last million years or so. These conditions for the free precession model to work are sufficiently unlikely that we cannot accept the model.

In all, we take the forced precession model to be the simplest explanation that explains all of our observations over the last twenty years.

## 9 Conclusions

We present 493 magnitudes on 246 nights from 1999 to 2006. This brings the Nereid photometry data base to a total of 774 magnitudes over 362 nights from 1987 to 2006. With this large data base from the last twenty years, Nereid has now one of the best photometric records of all outer Solar System bodies.

Our conclusions are:

(1) Nereid is continuing the year-to-year changes as seen by many groups in earlier years. There were no detectable variations around the phase curve from 1998-2000 and in 2006, but there were significant and substantial variations from 2003-2005. These year-to-year changes are now the central mystery of Nereid.

(2) Nereid is likely substantially nonspherical in shape. Based on the well-known shapes of similar-sized moons, we expect that Nereid has axial ratios of greater than 1.08:1 (i.e., >8% deviations from a sphere), while a significant fraction of comparable-sized moons have large axial ratios, perhaps typified by Hyperion with its 1.9:1 shape. The single 9x4 pixel image from *Voyager 2* puts little constraint on the shape of Nereid, as demonstrated by the inability to distinguish the round from the far-out-of-round moons in Figure 1.

(3) Nereid has an apparently stable phase curve of rather large amplitude (0.43 mag from 0° to 1.93°). This can only be explained as an opposition surge effect caused by coherent backscattering. We find that the effect has an amplitude of $B_{C0}$=0.50±0.05 and an angular width of $h_C$=0.7°±0.1°.

(4) Non-principal axis rotation will lead to free precession that can explain the observations. However, Nereid must be spherical to within ~0.05% to get a sufficiently long precession period and Nereid must have suffered a glancing impact from a >24 kilometer diameter body within the last million years so as to excite the non-principal axis rotation. This possibility does not seem reasonable.

(5) We propose that the year-to-year variations are caused by Nereid's pole precessing, such that in some years we have a pole pointing towards Earth and in other years we have the equatorial regions pointing towards Earth. A reasonable physical explanation is from forced precession, where Neptune exerts a tidal torque on an out-of-round Nereid. The precession period can be short enough to explain our data provided that Nereid has axis ratios of roughly 1.9:1 or greater. This is our preferred explanation.

We thank Alan Fitzsimmons and Robert T. J. McAteer for permission to include their R-band data from August 1999. We thank Tony Dobrovolskis for information on the rotational dynamics of Nereid. The National Aeronautics and Space Administration provided funds under grants NAG5-13533 and NAG5-13369.

**TABLE 1**
**Journal of Observations**

| Observing Dates | Julian Dates | Telescope | Filter | Number |
|---|---|---|---|---|
| 1999 June-Oct | 2451350 - 2451480 | CTIO 1.0m | V | 54 |
| 1999 Aug | 2451394 - 2451406 | La Palma 1.0m | R | 23 |
| 2000 July-Nov | 2451748 - 2451850 | CTIO 1.0m | V | 14 |
| 2003 Mar-Aug | 2452729 - 2452872 | CTIO 1.3m | BVI | 175 |
| 2004 July | 2453185 - 2453195 | CTIO 0.9m | V | 53 |
| 2004 Aug-Nov | 2453223 - 2453281 | CTIO 1.0m | V | 41 |
| 2005 May-Oct | 2453506 - 2453670 | CTIO 1.3m | V | 48 |
| 2005 June | 2453530 - 2453538 | CTIO 0.9m | V | 14 |
| 2006 May-Sep | 2453863 - 2454008 | CTIO 1.3m | V | 71 |

**TABLE 2**
**Nereid Magnitudes from 1999 to 2006**

| Julian Date | Band | $\phi$ (°) | Standard Magnitude |
|---|---|---|---|
| 2451350.745 | V | 1.09 | 19.46 ± 0.06 |
| 2451352.740 | V | 1.04 | 19.48 ± 0.05 |
| 2451354.747 | V | 0.98 | 19.40 ± 0.03 |
| 2451355.733 | V | 0.96 | 19.44 ± 0.03 |
| 2451365.723 | V | 0.66 | 19.31 ± 0.04 |
| 2451366.712 | V | 0.63 | 19.35 ± 0.04 |
| 2451367.708 | V | 0.60 | 19.30 ± 0.05 |
| 2451379.675 | V | 0.21 | 19.22 ± 0.03 |
| 2451380.675 | V | 0.18 | 19.20 ± 0.03 |
| 2451381.679 | V | 0.14 | 19.22 ± 0.03 |
| 2451382.678 | V | 0.11 | 19.17 ± 0.03 |
| 2451386.660 | V | 0.02 | 19.01 ± 0.05 |
| 2451389.679 | V | 0.12 | 19.28 ± 0.06 |
| 2451390.671 | V | 0.15 | 19.24 ± 0.08 |
| 2451391.701 | V | 0.19 | 19.34 ± 0.06 |
| 2451392.671 | V | 0.22 | 19.25 ± 0.04 |
| 2451393.669 | V | 0.25 | 19.26 ± 0.05 |
| 2451394.354 | R | 0.28 | 19.03 ± 0.12 |
| 2451394.423 | R | 0.28 | 18.91 ± 0.11 |
| 2451394.662 | V | 0.29 | 19.35 ± 0.04 |
| 2451395.330 | R | 0.31 | 18.84 ± 0.07 |
| 2451395.377 | R | 0.31 | 18.95 ± 0.08 |
| 2451396.404 | R | 0.34 | 18.89 ± 0.07 |
| 2451396.674 | V | 0.35 | 19.37 ± 0.05 |
| 2451397.362 | R | 0.37 | 19.03 ± 0.07 |
| 2451398.269 | R | 0.40 | 18.88 ± 0.08 |
| 2451398.328 | R | 0.41 | 18.95 ± 0.08 |
| 2451398.414 | R | 0.41 | 18.95 ± 0.08 |
| 2451399.266 | R | 0.44 | 18.81 ± 0.07 |
| 2451399.316 | R | 0.44 | 18.91 ± 0.07 |
| 2451399.390 | R | 0.44 | 18.84 ± 0.07 |
| 2451399.642 | V | 0.45 | 19.43 ± 0.04 |
| 2451401.248 | R | 0.50 | 18.88 ± 0.06 |
| 2451401.295 | R | 0.50 | 19.00 ± 0.06 |
| 2451401.648 | V | 0.51 | 19.41 ± 0.03 |
| 2451402.253 | R | 0.53 | 19.05 ± 0.07 |
| 2451402.333 | R | 0.53 | 18.95 ± 0.08 |
| 2451402.639 | V | 0.54 | 19.36 ± 0.03 |
| 2451403.258 | R | 0.56 | 18.92 ± 0.07 |
| 2451403.636 | V | 0.57 | 19.35 ± 0.04 |
| 2451404.256 | R | 0.59 | 18.91 ± 0.06 |
| 2451404.307 | R | 0.60 | 18.87 ± 0.08 |
| 2451405.241 | R | 0.63 | 18.88 ± 0.11 |
| 2451405.317 | R | 0.63 | 18.84 ± 0.08 |
| 2451406.259 | R | 0.66 | 18.85 ± 0.10 |
| 2451406.313 | R | 0.66 | 18.99 ± 0.09 |
| 2451406.640 | V | 0.67 | 19.43 ± 0.06 |
| 2451407.624 | V | 0.70 | 19.41 ± 0.03 |
| 2451408.631 | V | 0.73 | 19.42 ± 0.04 |
| 2451410.616 | V | 0.79 | 19.44 ± 0.04 |
| 2451412.618 | V | 0.85 | 19.42 ± 0.06 |
| 2451432.569 | V | 1.38 | 19.44 ± 0.04 |
| 2451433.539 | V | 1.40 | 19.46 ± 0.04 |
| 2451441.559 | V | 1.56 | 19.51 ± 0.09 |
| 2451443.509 | V | 1.60 | 19.71 ± 0.15 |
| 2451444.499 | V | 1.61 | 19.37 ± 0.09 |
| 2451445.499 | V | 1.63 | 19.54 ± 0.07 |
| 2451446.499 | V | 1.65 | 19.47 ± 0.10 |
| 2451449.489 | V | 1.69 | 19.52 ± 0.05 |
| 2451451.489 | V | 1.72 | 19.56 ± 0.08 |
| 2451454.478 | V | 1.76 | 19.49 ± 0.04 |
| 2451455.478 | V | 1.77 | 19.57 ± 0.05 |
| 2451458.458 | V | 1.80 | 19.54 ± 0.04 |
| 2451460.488 | V | 1.82 | 19.60 ± 0.05 |
| 2451461.498 | V | 1.83 | 19.54 ± 0.04 |
| 2451463.418 | V | 1.85 | 19.53 ± 0.04 |
| 2451465.447 | V | 1.86 | 19.55 ± 0.04 |
| 2451466.507 | V | 1.87 | 19.53 ± 0.05 |
| 2451467.437 | V | 1.87 | 19.54 ± 0.05 |
| 2451468.497 | V | 1.88 | 19.55 ± 0.10 |
| 2451471.407 | V | 1.89 | 19.54 ± 0.06 |
| 2451472.407 | V | 1.89 | 19.53 ± 0.07 |
| 2451475.406 | V | 1.89 | 19.48 ± 0.06 |
| 2451477.396 | V | 1.89 | 19.57 ± 0.05 |
| 2451478.416 | V | 1.89 | 19.55 ± 0.05 |
| 2451479.406 | V | 1.89 | 19.56 ± 0.04 |
| 2451480.406 | V | 1.89 | 19.58 ± 0.04 |
| 2451748.669 | V | 0.16 | 19.24 ± 0.04 |
| 2451750.665 | V | 0.10 | 19.18 ± 0.03 |
| 2451752.667 | V | 0.03 | 19.13 ± 0.03 |
| 2451754.653 | V | 0.04 | 19.16 ± 0.03 |
| 2451756.673 | V | 0.10 | 19.23 ± 0.04 |
| 2451758.587 | V | 0.17 | 19.24 ± 0.03 |
| 2451766.628 | V | 0.43 | 19.30 ± 0.04 |
| 2451768.632 | V | 0.49 | 19.42 ± 0.06 |
| 2451782.583 | V | 0.92 | 19.43 ± 0.04 |
| 2451784.575 | V | 0.98 | 19.40 ± 0.04 |
| 2451808.518 | V | 1.55 | 19.48 ± 0.05 |
| 2451810.510 | V | 1.59 | 19.55 ± 0.03 |
| 2451819.425 | V | 1.73 | 19.61 ± 0.05 |
| 2451850.392 | V | 1.88 | 19.53 ± 0.04 |
| 2452729.712 | B | 1.61 | 19.84 ± 0.12 |
| 2452731.730 | B | 1.64 | 19.82 ± 0.12 |
| 2452731.732 | I | 1.64 | 18.75 ± 0.19 |
| 2452751.708 | V | 1.88 | 19.58 ± 0.20 |
| 2452751.712 | B | 1.88 | 20.36 ± 0.17 |
| 2452755.715 | V | 1.90 | 19.95 ± 0.14 |
| 2452755.716 | V | 1.90 | 19.82 ± 0.12 |
| 2452755.719 | B | 1.90 | 20.42 ± 0.10 |
| 2452757.720 | V | 1.91 | 19.96 ± 0.16 |
| 2452757.721 | V | 1.91 | 20.05 ± 0.16 |
| 2452757.723 | B | 1.91 | 20.50 ± 0.07 |
| 2452759.737 | V | 1.92 | 19.91 ± 0.10 |
| 2452759.741 | B | 1.92 | 20.37 ± 0.06 |
| 2452763.603 | V | 1.92 | 19.84 ± 0.12 |
| 2452763.604 | V | 1.92 | 19.67 ± 0.10 |
| 2452763.607 | B | 1.92 | 20.48 ± 0.08 |
| 2452765.658 | V | 1.92 | 19.67 ± 0.12 |
| 2452765.662 | B | 1.92 | 20.21 ± 0.06 |
| 2452765.664 | I | 1.92 | 18.83 ± 0.19 |
| 2452771.621 | V | 1.91 | 19.38 ± 0.14 |
| 2452771.622 | V | 1.91 | 19.25 ± 0.12 |
| 2452771.625 | B | 1.91 | 19.72 ± 0.12 |
| 2452778.685 | B | 1.87 | 19.72 ± 0.17 |
| 2452787.692 | V | 1.77 | 19.35 ± 0.09 |
| 2452787.693 | V | 1.77 | 19.46 ± 0.08 |
| 2452787.696 | B | 1.77 | 20.12 ± 0.05 |
| 2452787.698 | I | 1.77 | 18.87 ± 0.16 |
| 2452789.687 | V | 1.75 | 19.86 ± 0.14 |
| 2452789.688 | V | 1.75 | 19.42 ± 0.10 |
| 2452789.690 | B | 1.75 | 20.01 ± 0.06 |
| 2452789.693 | I | 1.75 | 18.65 ± 0.19 |
| 2452791.685 | V | 1.72 | 19.64 ± 0.10 |
| 2452791.686 | V | 1.72 | 19.67 ± 0.10 |
| 2452791.688 | B | 1.72 | 20.11 ± 0.11 |
| 2452791.691 | I | 1.72 | 18.65 ± 0.14 |
| 2452793.729 | V | 1.69 | 19.54 ± 0.20 |
| 2452793.730 | V | 1.69 | 19.51 ± 0.16 |
| 2452793.733 | B | 1.69 | 19.99 ± 0.08 |
| 2452817.726 | V | 1.18 | 19.44 ± 0.10 |
| 2452817.728 | V | 1.18 | 19.43 ± 0.11 |
| 2452817.730 | B | 1.18 | 20.16 ± 0.07 |
| 2452819.698 | V | 1.13 | 19.51 ± 0.09 |
| 2452819.699 | V | 1.13 | 19.59 ± 0.09 |
| 2452819.701 | B | 1.13 | 20.21 ± 0.05 |
| 2452819.704 | I | 1.13 | 18.69 ± 0.16 |
| 2452821.687 | V | 1.08 | 19.27 ± 0.20 |
| 2452821.688 | V | 1.08 | 19.33 ± 0.17 |
| 2452821.691 | B | 1.08 | 19.94 ± 0.11 |
| 2452829.716 | V | 0.85 | 19.30 ± 0.09 |
| 2452829.718 | V | 0.85 | 19.47 ± 0.11 |
| 2452829.720 | B | 0.85 | 20.00 ± 0.07 |
| 2452829.722 | I | 0.85 | 18.39 ± 0.19 |
| 2452831.651 | V | 0.79 | 19.46 ± 0.15 |
| 2452831.652 | V | 0.79 | 19.37 ± 0.14 |
| 2452831.654 | B | 0.79 | 20.22 ± 0.13 |
| 2452831.657 | I | 0.79 | 18.76 ± 0.20 |
| 2452832.589 | V | 0.76 | 19.49 ± 0.17 |
| 2452832.591 | V | 0.76 | 19.29 ± 0.15 |
| 2452832.593 | B | 0.76 | 20.05 ± 0.14 |
| 2452832.595 | I | 0.76 | 18.68 ± 0.17 |
| 2452837.652 | B | 0.61 | 19.41 ± 0.19 |
| 2452839.642 | V | 0.54 | 19.31 ± 0.12 |
| 2452839.644 | V | 0.54 | 19.35 ± 0.14 |
| 2452839.646 | B | 0.54 | 20.07 ± 0.12 |
| 2452839.649 | I | 0.54 | 18.62 ± 0.19 |
| 2452841.692 | V | 0.48 | 19.15 ± 0.12 |
| 2452841.693 | V | 0.48 | 19.10 ± 0.10 |
| 2452841.695 | B | 0.48 | 19.81 ± 0.08 |
| 2452841.698 | I | 0.48 | 18.40 ± 0.16 |
| 2452843.633 | V | 0.41 | 19.40 ± 0.08 |
| 2452843.635 | V | 0.41 | 19.39 ± 0.08 |
| 2452843.637 | B | 0.41 | 20.06 ± 0.06 |
| 2452843.639 | I | 0.41 | 18.80 ± 0.17 |
| 2452844.620 | V | 0.38 | 19.17 ± 0.07 |
| 2452844.621 | V | 0.38 | 19.37 ± 0.08 |
| 2452844.624 | B | 0.38 | 20.02 ± 0.05 |
| 2452844.626 | I | 0.38 | 18.64 ± 0.13 |
| 2452845.635 | V | 0.35 | 19.31 ± 0.07 |
| 2452845.636 | V | 0.35 | 19.45 ± 0.09 |
| 2452845.638 | B | 0.35 | 19.83 ± 0.05 |
| 2452845.641 | I | 0.35 | 18.41 ± 0.12 |
| 2452846.645 | V | 0.32 | 19.33 ± 0.16 |
| 2452846.647 | V | 0.32 | 19.37 ± 0.19 |
| 2452846.649 | B | 0.32 | 19.90 ± 0.11 |
| 2452847.636 | V | 0.28 | 19.32 ± 0.11 |
| 2452847.638 | V | 0.28 | 19.13 ± 0.10 |
| 2452847.640 | B | 0.28 | 19.82 ± 0.07 |
| 2452847.642 | I | 0.28 | 18.61 ± 0.19 |
| 2452851.673 | V | 0.15 | 19.24 ± 0.04 |
| 2452851.674 | V | 0.15 | 19.23 ± 0.05 |
| 2452851.677 | B | 0.15 | 19.80 ± 0.04 |
| 2452851.679 | I | 0.15 | 18.44 ± 0.09 |
| 2452852.670 | V | 0.12 | 19.18 ± 0.04 |
| 2452852.671 | V | 0.12 | 19.19 ± 0.04 |
| 2452852.674 | B | 0.12 | 19.86 ± 0.03 |
| 2452852.676 | I | 0.12 | 18.37 ± 0.07 |
| 2452853.655 | V | 0.08 | 19.14 ± 0.04 |
| 2452853.656 | V | 0.08 | 19.15 ± 0.04 |
| 2452853.659 | B | 0.08 | 19.87 ± 0.02 |
| 2452853.661 | I | 0.08 | 18.42 ± 0.07 |
| 2452854.604 | V | 0.05 | 19.09 ± 0.04 |
| 2452854.605 | V | 0.05 | 19.13 ± 0.04 |
| 2452854.610 | I | 0.05 | 18.36 ± 0.06 |
| 2452855.371 | V | 0.03 | 19.06 ± 0.08 |
| 2452855.373 | V | 0.03 | 19.09 ± 0.08 |
| 2452855.375 | B | 0.03 | 19.80 ± 0.06 |
| 2452855.378 | I | 0.03 | 18.39 ± 0.12 |
| 2452856.514 | V | 0.01 | 19.06 ± 0.06 |
| 2452856.515 | V | 0.01 | 19.14 ± 0.07 |
| 2452856.518 | B | 0.01 | 19.75 ± 0.05 |
| 2452856.520 | I | 0.01 | 18.35 ± 0.10 |
| 2452856.677 | V | 0.02 | 19.15 ± 0.09 |
| 2452856.678 | V | 0.02 | 19.02 ± 0.08 |
| 2452856.681 | B | 0.02 | 19.82 ± 0.05 |
| 2452856.683 | I | 0.02 | 18.44 ± 0.12 |
| 2452857.399 | V | 0.04 | 19.08 ± 0.05 |
| 2452857.402 | B | 0.04 | 19.74 ± 0.04 |
| 2452857.404 | I | 0.04 | 18.40 ± 0.06 |
| 2452857.437 | V | 0.04 | 19.43 ± 0.06 |
| 2452857.438 | V | 0.04 | 19.41 ± 0.05 |
| 2452857.441 | B | 0.04 | 19.83 ± 0.03 |
| 2452857.443 | I | 0.04 | 18.61 ± 0.07 |
| 2452857.482 | V | 0.04 | 19.15 ± 0.04 |
| 2452857.483 | V | 0.04 | 19.22 ± 0.05 |
| 2452857.485 | B | 0.04 | 19.76 ± 0.03 |
| 2452857.488 | I | 0.04 | 18.51 ± 0.07 |
| 2452857.676 | V | 0.05 | 19.22 ± 0.05 |
| 2452857.677 | V | 0.05 | 19.16 ± 0.05 |
| 2452857.679 | B | 0.05 | 19.81 ± 0.03 |
| 2452857.682 | I | 0.05 | 18.44 ± 0.09 |
| 2452858.372 | V | 0.07 | 19.02 ± 0.05 |
| 2452858.374 | V | 0.07 | 19.17 ± 0.05 |
| 2452858.376 | B | 0.07 | 19.79 ± 0.04 |
| 2452858.378 | I | 0.07 | 18.46 ± 0.07 |
| 2452858.505 | V | 0.08 | 19.24 ± 0.04 |
| 2452858.507 | V | 0.08 | 19.19 ± 0.04 |
| 2452858.509 | B | 0.08 | 19.83 ± 0.03 |
| 2452858.511 | I | 0.08 | 18.62 ± 0.06 |
| 2452858.650 | V | 0.08 | 19.21 ± 0.04 |
| 2452858.651 | V | 0.08 | 19.23 ± 0.04 |
| 2452858.654 | B | 0.08 | 19.82 ± 0.02 |
| 2452858.656 | I | 0.08 | 18.58 ± 0.08 |
| 2452859.633 | V | 0.11 | 19.20 ± 0.07 |
| 2452859.634 | V | 0.11 | 19.16 ± 0.07 |
| 2452859.636 | B | 0.11 | 19.71 ± 0.05 |
| 2452859.639 | I | 0.11 | 18.43 ± 0.09 |
| 2452860.621 | V | 0.15 | 19.10 ± 0.06 |
| 2452860.622 | V | 0.15 | 19.18 ± 0.07 |
| 2452860.624 | B | 0.15 | 19.95 ± 0.06 |
| 2452860.627 | I | 0.15 | 18.43 ± 0.07 |
| 2452862.605 | V | 0.21 | 19.17 ± 0.17 |
| 2452862.606 | V | 0.21 | 19.11 ± 0.16 |
| 2452862.608 | B | 0.21 | 19.80 ± 0.12 |
| 2452865.619 | B | 0.31 | 19.52 ± 0.13 |
| 2452866.575 | V | 0.34 | 19.22 ± 0.12 |
| 2452866.576 | V | 0.34 | 19.40 ± 0.12 |
| 2452866.579 | B | 0.34 | 19.97 ± 0.11 |
| 2452866.581 | I | 0.34 | 18.40 ± 0.15 |
| 2452867.559 | V | 0.37 | 19.32 ± 0.07 |
| 2452867.561 | V | 0.37 | 19.27 ± 0.06 |
| 2452867.563 | B | 0.37 | 20.01 ± 0.05 |
| 2452867.565 | I | 0.37 | 18.65 ± 0.08 |
| 2452868.538 | V | 0.41 | 19.39 ± 0.05 |
| 2452868.539 | V | 0.41 | 19.45 ± 0.05 |
| 2452868.541 | B | 0.41 | 20.01 ± 0.04 |
| 2452868.544 | I | 0.41 | 18.70 ± 0.07 |
| 2452869.545 | V | 0.44 | 19.38 ± 0.08 |
| 2452869.546 | V | 0.44 | 19.35 ± 0.06 |
| 2452869.549 | B | 0.44 | 20.08 ± 0.04 |
| 2452869.551 | I | 0.44 | 18.47 ± 0.11 |
| 2452871.567 | V | 0.50 | 19.30 ± 0.06 |
| 2452871.569 | V | 0.50 | 19.34 ± 0.09 |
| 2452871.571 | B | 0.50 | 19.83 ± 0.05 |
| 2452871.573 | I | 0.50 | 18.68 ± 0.14 |
| 2452872.530 | B | 0.53 | 20.29 ± 0.18 |
| 2453185.734 | V | 1.17 | 19.17 ± 0.04 |
| 2453185.746 | V | 1.17 | 19.14 ± 0.04 |
| 2453185.757 | V | 1.17 | 19.12 ± 0.04 |
| 2453186.591 | V | 1.15 | 19.20 ± 0.05 |
| 2453186.733 | V | 1.15 | 19.28 ± 0.04 |
| 2453186.744 | V | 1.15 | 19.29 ± 0.04 |
| 2453186.755 | V | 1.15 | 19.36 ± 0.04 |
| 2453186.766 | V | 1.14 | 19.36 ± 0.04 |
| 2453186.777 | V | 1.14 | 19.22 ± 0.05 |
| 2453187.712 | V | 1.12 | 19.32 ± 0.07 |
| 2453187.717 | V | 1.12 | 19.41 ± 0.09 |
| 2453187.724 | V | 1.12 | 19.30 ± 0.12 |
| 2453187.726 | V | 1.12 | 19.22 ± 0.10 |
| 2453188.699 | V | 1.09 | 19.34 ± 0.07 |
| 2453188.710 | V | 1.09 | 19.41 ± 0.07 |
| 2453188.721 | V | 1.09 | 19.36 ± 0.07 |
| 2453188.732 | V | 1.09 | 19.24 ± 0.07 |
| 2453188.743 | V | 1.09 | 19.29 ± 0.07 |
| 2453188.754 | V | 1.09 | 19.51 ± 0.09 |
| 2453188.765 | V | 1.09 | 19.52 ± 0.10 |
| 2453189.525 | V | 1.07 | 19.34 ± 0.09 |
| 2453189.537 | V | 1.07 | 19.38 ± 0.09 |
| 2453189.632 | V | 1.07 | 19.47 ± 0.12 |
| 2453192.677 | V | 0.98 | 19.39 ± 0.08 |
| 2453192.691 | V | 0.98 | 19.30 ± 0.06 |
| 2453192.702 | V | 0.98 | 19.22 ± 0.06 |
| 2453192.713 | V | 0.98 | 19.18 ± 0.06 |
| 2453192.724 | V | 0.98 | 19.30 ± 0.07 |
| 2453192.734 | V | 0.98 | 19.32 ± 0.07 |
| 2453192.745 | V | 0.98 | 19.25 ± 0.07 |
| 2453192.756 | V | 0.98 | 19.19 ± 0.07 |
| 2453192.767 | V | 0.98 | 19.23 ± 0.08 |
| 2453193.587 | V | 0.96 | 19.18 ± 0.07 |
| 2453193.599 | V | 0.96 | 19.11 ± 0.06 |
| 2453193.770 | V | 0.95 | 19.14 ± 0.07 |
| 2453193.781 | V | 0.95 | 19.30 ± 0.12 |
| 2453194.520 | V | 0.93 | 19.19 ± 0.05 |
| 2453194.531 | V | 0.93 | 19.25 ± 0.04 |
| 2453194.542 | V | 0.93 | 19.23 ± 0.04 |
| 2453194.553 | V | 0.93 | 19.39 ± 0.05 |
| 2453194.564 | V | 0.93 | 19.28 ± 0.04 |
| 2453194.580 | V | 0.93 | 19.37 ± 0.12 |
| 2453194.582 | V | 0.93 | 19.21 ± 0.04 |
| 2453194.592 | V | 0.93 | 19.28 ± 0.04 |
| 2453194.603 | V | 0.93 | 19.28 ± 0.04 |
| 2453195.697 | V | 0.90 | 19.24 ± 0.07 |
| 2453195.709 | V | 0.89 | 19.25 ± 0.05 |
| 2453195.720 | V | 0.89 | 19.30 ± 0.05 |
| 2453195.731 | V | 0.89 | 19.27 ± 0.04 |
| 2453195.742 | V | 0.89 | 19.25 ± 0.05 |
| 2453195.752 | V | 0.89 | 19.18 ± 0.05 |
| 2453195.763 | V | 0.89 | 19.27 ± 0.06 |
| 2453195.774 | V | 0.89 | 19.24 ± 0.06 |
| 2453223.626 | V | 0.00 | 18.75 ± 0.06 |
| 2453223.672 | V | 0.00 | 18.78 ± 0.06 |
| 2453224.600 | V | 0.03 | 18.84 ± 0.06 |
| 2453225.593 | V | 0.06 | 18.71 ± 0.05 |
| 2453229.563 | V | 0.19 | 19.04 ± 0.04 |
| 2453229.598 | V | 0.19 | 19.07 ± 0.05 |
| 2453230.625 | V | 0.23 | 19.22 ± 0.05 |
| 2453237.558 | V | 0.45 | 19.28 ± 0.04 |
| 2453238.562 | V | 0.49 | 19.18 ± 0.04 |
| 2453239.560 | V | 0.52 | 19.24 ± 0.05 |
| 2453241.560 | V | 0.58 | 18.99 ± 0.06 |
| 2453243.537 | V | 0.64 | 19.29 ± 0.08 |
| 2453248.548 | V | 0.80 | 18.93 ± 0.06 |
| 2453248.589 | V | 0.80 | 19.15 ± 0.07 |
| 2453249.497 | V | 0.83 | 19.06 ± 0.06 |
| 2453250.542 | V | 0.86 | 19.10 ± 0.05 |
| 2453251.538 | V | 0.89 | 19.18 ± 0.05 |
| 2453251.586 | V | 0.89 | 19.09 ± 0.05 |
| 2453254.552 | V | 0.97 | 19.09 ± 0.06 |
| 2453258.559 | V | 1.09 | 19.27 ± 0.04 |
| 2453259.524 | V | 1.11 | 19.20 ± 0.04 |
| 2453260.481 | V | 1.14 | 19.21 ± 0.06 |
| 2453261.468 | V | 1.16 | 19.20 ± 0.05 |
| 2453262.518 | V | 1.19 | 19.17 ± 0.04 |
| 2453263.466 | V | 1.21 | 19.27 ± 0.04 |
| 2453264.379 | V | 1.24 | 19.28 ± 0.07 |
| 2453265.409 | V | 1.26 | 19.26 ± 0.04 |
| 2453269.518 | V | 1.36 | 19.27 ± 0.08 |
| 2453271.519 | V | 1.41 | 19.36 ± 0.06 |
| 2453274.462 | V | 1.47 | 19.50 ± 0.09 |
| 2453275.445 | V | 1.49 | 19.29 ± 0.07 |
| 2453275.498 | V | 1.49 | 19.18 ± 0.09 |
| 2453277.452 | V | 1.53 | 19.42 ± 0.08 |
| 2453277.503 | V | 1.53 | 19.35 ± 0.07 |
| 2453278.483 | V | 1.55 | 19.23 ± 0.07 |
| 2453278.518 | V | 1.55 | 19.12 ± 0.06 |
| 2453279.443 | V | 1.57 | 19.17 ± 0.07 |
| 2453279.501 | V | 1.57 | 19.10 ± 0.06 |
| 2453280.366 | V | 1.58 | 19.10 ± 0.04 |
| 2453281.450 | V | 1.60 | 19.25 ± 0.05 |
| 2453281.484 | V | 1.60 | 19.11 ± 0.05 |
| 2453506.691 | V | 1.91 | 19.57 ± 0.02 |
| 2453508.707 | V | 1.90 | 19.57 ± 0.02 |
| 2453511.694 | V | 1.88 | 19.53 ± 0.02 |
| 2453515.689 | V | 1.85 | 19.57 ± 0.04 |
| 2453517.696 | V | 1.83 | 19.38 ± 0.05 |
| 2453521.684 | V | 1.79 | 19.46 ± 0.05 |
| 2453523.675 | V | 1.77 | 19.45 ± 0.15 |
| 2453525.668 | V | 1.74 | 19.28 ± 0.15 |
| 2453527.665 | V | 1.71 | 19.21 ± 0.14 |
| 2453529.656 | V | 1.68 | 19.39 ± 0.16 |
| 2453530.696 | V | 1.66 | 19.47 ± 0.04 |
| 2453530.707 | V | 1.66 | 19.47 ± 0.03 |
| 2453530.719 | V | 1.66 | 19.59 ± 0.08 |
| 2453530.730 | V | 1.66 | 19.48 ± 0.03 |
| 2453530.743 | V | 1.66 | 19.47 ± 0.06 |
| 2453530.754 | V | 1.66 | 19.43 ± 0.09 |
| 2453538.588 | V | 1.52 | 19.54 ± 0.02 |
| 2453538.600 | V | 1.52 | 19.59 ± 0.02 |
| 2453538.629 | V | 1.52 | 19.63 ± 0.02 |
| 2453538.640 | V | 1.51 | 19.56 ± 0.02 |
| 2453538.669 | V | 1.51 | 19.49 ± 0.02 |
| 2453538.679 | V | 1.51 | 19.51 ± 0.02 |
| 2453538.714 | V | 1.51 | 19.51 ± 0.02 |
| 2453538.757 | V | 1.51 | 19.55 ± 0.02 |
| 2453547.682 | V | 1.31 | 19.57 ± 0.05 |
| 2453550.682 | V | 1.24 | 19.54 ± 0.03 |
| 2453575.619 | V | 0.52 | 19.37 ± 0.04 |
| 2453577.594 | V | 0.45 | 19.29 ± 0.02 |
| 2453586.424 | V | 0.16 | 19.24 ± 0.02 |
| 2453586.632 | V | 0.16 | 19.24 ± 0.02 |
| 2453587.674 | V | 0.12 | 19.07 ± 0.10 |
| 2453588.373 | V | 0.10 | 18.82 ± 0.11 |
| 2453588.658 | V | 0.09 | 19.07 ± 0.08 |
| 2453589.370 | V | 0.06 | 19.18 ± 0.02 |
| 2453589.689 | V | 0.05 | 19.18 ± 0.02 |
| 2453590.367 | V | 0.03 | 19.12 ± 0.02 |
| 2453590.683 | V | 0.02 | 19.02 ± 0.02 |
| 2453591.387 | V | 0.01 | 19.13 ± 0.02 |
| 2453591.657 | V | 0.01 | 19.11 ± 0.02 |
| 2453592.465 | V | 0.04 | 19.20 ± 0.02 |
| 2453592.672 | V | 0.05 | 19.17 ± 0.02 |
| 2453593.506 | V | 0.07 | 19.23 ± 0.02 |
| 2453593.654 | V | 0.08 | 19.23 ± 0.02 |
| 2453594.400 | V | 0.10 | 19.26 ± 0.02 |
| 2453594.641 | V | 0.11 | 19.20 ± 0.04 |
| 2453603.634 | V | 0.41 | 19.34 ± 0.05 |
| 2453607.576 | V | 0.53 | 19.38 ± 0.02 |
| 2453619.549 | V | 0.90 | 19.43 ± 0.02 |
| 2453621.505 | V | 0.96 | 19.50 ± 0.02 |
| 2453626.470 | V | 1.10 | 19.24 ± 0.13 |
| 2453629.527 | V | 1.18 | 19.33 ± 0.13 |
| 2453631.447 | V | 1.23 | 19.46 ± 0.05 |
| 2453633.489 | V | 1.28 | 19.44 ± 0.10 |
| 2453635.479 | V | 1.32 | 19.44 ± 0.05 |
| 2453639.438 | V | 1.42 | 19.53 ± 0.02 |
| 2453641.444 | V | 1.45 | 19.51 ± 0.02 |
| 2453642.499 | V | 1.48 | 19.50 ± 0.03 |
| 2453644.439 | V | 1.52 | 19.49 ± 0.07 |
| 2453662.385 | V | 1.79 | 19.46 ± 0.07 |
| 2453665.397 | V | 1.82 | 19.59 ± 0.03 |
| 2453668.367 | V | 1.85 | 19.49 ± 0.02 |
| 2453670.348 | V | 1.86 | 19.23 ± 0.02 |
| 2453863.633 | V | 1.92 | 19.53 ± 0.04 |
| 2453865.719 | V | 1.93 | 19.59 ± 0.03 |
| 2453867.687 | V | 1.93 | 19.49 ± 0.04 |
| 2453868.739 | V | 1.93 | 19.44 ± 0.13 |
| 2453871.694 | V | 1.92 | 19.42 ± 0.05 |
| 2453873.761 | V | 1.91 | 19.48 ± 0.05 |
| 2453877.714 | V | 1.89 | 19.59 ± 0.03 |
| 2453891.545 | V | 1.76 | 19.41 ± 0.08 |
| 2453891.555 | V | 1.76 | 19.44 ± 0.05 |
| 2453891.559 | V | 1.76 | 19.50 ± 0.04 |
| 2453891.563 | V | 1.76 | 19.49 ± 0.05 |
| 2453891.571 | V | 1.76 | 19.58 ± 0.04 |
| 2453891.584 | V | 1.76 | 19.54 ± 0.04 |
| 2453891.600 | V | 1.76 | 19.55 ± 0.04 |
| 2453891.604 | V | 1.76 | 19.54 ± 0.04 |
| 2453891.618 | V | 1.76 | 19.54 ± 0.03 |
| 2453891.665 | V | 1.76 | 19.54 ± 0.03 |
| 2453891.669 | V | 1.76 | 19.58 ± 0.03 |
| 2453891.683 | V | 1.76 | 19.54 ± 0.03 |
| 2453891.695 | V | 1.76 | 19.61 ± 0.04 |
| 2453891.699 | V | 1.76 | 19.58 ± 0.05 |
| 2453891.703 | V | 1.76 | 19.53 ± 0.06 |
| 2453891.714 | V | 1.76 | 19.57 ± 0.08 |
| 2453891.718 | V | 1.76 | 19.56 ± 0.07 |
| 2453891.722 | V | 1.76 | 19.52 ± 0.08 |
| 2453891.726 | V | 1.76 | 19.53 ± 0.05 |
| 2453891.737 | V | 1.76 | 19.50 ± 0.10 |
| 2453891.741 | V | 1.76 | 19.59 ± 0.20 |
| 2453900.689 | V | 1.62 | 19.47 ± 0.08 |
| 2453903.628 | V | 1.57 | 19.52 ± 0.10 |
| 2453907.659 | V | 1.49 | 19.52 ± 0.02 |
| 2453908.639 | V | 1.46 | 19.53 ± 0.02 |
| 2453911.684 | V | 1.40 | 19.51 ± 0.03 |
| 2453932.681 | V | 0.84 | 19.38 ± 0.16 |
| 2453934.603 | V | 0.78 | 19.45 ± 0.02 |
| 2453936.503 | V | 0.72 | 19.47 ± 0.02 |
| 2453937.575 | V | 0.69 | 19.42 ± 0.02 |
| 2453944.584 | V | 0.47 | 19.39 ± 0.03 |
| 2453946.445 | V | 0.41 | 19.36 ± 0.02 |
| 2453959.576 | V | 0.03 | 19.19 ± 0.05 |
| 2453960.571 | V | 0.06 | 19.16 ± 0.03 |
| 2453960.614 | V | 0.06 | 19.20 ± 0.03 |
| 2453961.516 | V | 0.09 | 19.22 ± 0.03 |
| 2453961.553 | V | 0.09 | 19.21 ± 0.03 |
| 2453961.602 | V | 0.09 | 19.21 ± 0.03 |
| 2453963.585 | V | 0.16 | 19.27 ± 0.02 |
| 2453963.624 | V | 0.16 | 19.23 ± 0.02 |
| 2453964.556 | V | 0.19 | 19.26 ± 0.02 |
| 2453964.615 | V | 0.19 | 19.29 ± 0.02 |
| 2453964.615 | V | 0.19 | 19.29 ± 0.02 |
| 2453965.418 | V | 0.22 | 19.25 ± 0.02 |
| 2453965.649 | V | 0.23 | 19.33 ± 0.02 |
| 2453966.639 | V | 0.26 | 19.33 ± 0.02 |
| 2453967.594 | V | 0.29 | 19.34 ± 0.02 |
| 2453967.643 | V | 0.29 | 19.36 ± 0.02 |
| 2453968.632 | V | 0.32 | 19.35 ± 0.03 |
| 2453969.609 | V | 0.36 | 19.26 ± 0.03 |
| 2453970.620 | V | 0.39 | 19.43 ± 0.03 |
| 2453971.606 | V | 0.42 | 19.33 ± 0.03 |
| 2453973.579 | V | 0.48 | 19.42 ± 0.02 |
| 2453974.563 | V | 0.52 | 19.40 ± 0.02 |
| 2453975.552 | V | 0.55 | 19.42 ± 0.02 |
| 2453976.567 | V | 0.58 | 19.40 ± 0.02 |
| 2453979.564 | V | 0.67 | 19.38 ± 0.03 |
| 2453989.522 | V | 0.97 | 19.39 ± 0.06 |
| 2453991.430 | V | 1.03 | 19.48 ± 0.02 |
| 2453994.503 | V | 1.11 | 19.45 ± 0.02 |
| 2453996.507 | V | 1.16 | 19.48 ± 0.02 |
| 2454002.372 | V | 1.31 | 19.41 ± 0.04 |
| 2454004.428 | V | 1.36 | 19.50 ± 0.02 |
| 2454008.420 | V | 1.45 | 19.49 ± 0.02 |

**TABLE 3**
**Average phase function from 1998, 1999, 2000, and 2006**

| $\langle\phi\rangle$ | $\phi$ range | Number | $\langle m-5*Log_{10}[r\Delta/900]\rangle$ | Slope (mag/°) |
|---|---|---|---|---|
| 0.05° | 0.02°-0.09° | 16 | 19.18±0.01 | |
| | | | | 0.55±0.07 |
| 0.25° | 0.20°-0.30° | 13 | 19.29±0.01 | |
| | | | | 0.36±0.06 |
| 0.50° | 0.40°-0.60° | 19 | 19.38±0.01 | |
| | | | | 0.16±0.03 |
| 1.00° | 0.90°-1.10° | 16 | 19.46±0.01 | |
| | | | | 0.12±0.03 |
| 1.50° | 1.35-1.65° | 19 | 19.52±0.01 | |
| | | | | 0.10±0.03 |
| 1.90° | 1.85°-1.93° | 28 | 19.56±0.01 | |

**TABLE 4**
**Variations about best-fit Hapke phase function**

| Years | $\chi^2/N_{DoF}$ | RMS | Offset | Variations about Hapke phase curve |
|---|---|---|---|---|
| 1987-90 | 1464/26 | 0.40 | 0.07 | Large amplitude variations (both brightening and fading) |
| 1993-97 | 2505/49 | 0.07 | 0.00 | Moderate amplitude variations |
| 1998 | 93/57 | 0.06 | 0.01 | No variability |
| 1999 | 68/54 | 0.06 | -0.02 | No variability |
| 2000 | 12/14 | 0.04 | -0.01 | No variability |
| 2003 | 205/88 | 0.14 | 0.00 | Moderate amplitude variations (moderate error bars) |
| 2004 | 2192/94 | 0.11 | -0.24 | Moderate amplitude variations centered 0.24 mag brighter |
| 2005 | 481/62 | 0.10 | -0.06 | Episodes of ~0.2 mag brightening |
| 2006 | 147/71 | 0.05 | -0.01 | No variability |

**TABLE 5**
**Conservative amplitude limits**

| Years | 3-sigma amplitude limit (mag) 0.1-50 days | 0.240±0.003 days |
|---|---|---|
| 1998 | 0.12 | 0.10 |
| 1999 | 0.11 | 0.09 |
| 2000 | 0.14 | 0.11 |
| 2003 | 0.23 | 0.19 |
| 2004 | 0.19 | 0.15 |
| 2005 | 0.17 | 0.14 |
| 2006 | 0.09 | 0.07 |
| 1998-2000 | 0.08 | 0.07 |

**TABLE 6**
**Moons in the mosaic (Fig. 1)**

| Panel | Moon | Planet | Radius (km) | Shape | Source[1] |
|---|---|---|---|---|---|
| I | Moon | Earth | 1738 | Round with ~0.1% relief | … |
| B | Iapetus | Saturn | 730 | Round to 5%, 3% relief | PIA07660 |
| G | Proteus | Neptune | 218 x 208 x 201 | Box-shaped, ~20% structure | … |
| C | Hyperion | Saturn | 205 x 130 x 110 | 1.9:1 | PIA06645 |
| A | Mimas | Saturn | 196 | Round to 8%, 8% relief | PIA06582 |
| D | Nereid | Neptune | 170±25 | ? | PIA00054 |
| H | Phoebe | Saturn | 110 | Round to 9%, 7% relief | PIA06062 |
| E | Pandora | Saturn | 52 x 40 x 30 | 1.7:1 | PIA07632 |
| F | Calypso | Saturn | 15 x 11 x 7 | 2.1:1 | PIA07633 |

[1] NASA PIA images can be found at http://photojournal.jpl.nasa.gov.

**TABLE 7**
**Shapes of Nereid-sized moons**

| Moon | Planet | Radius (km) | Shape |
|---|---|---|---|
| Proteus | Neptune | 218 x 208 x 201 | Box-shaped, ~20% structure |
| Hyperion | Saturn | 205 x 130 x 110 | 1.9:1 |
| Mimas | Saturn | 196 | Round to 8%, 8% relief |
| Nereid | Neptune | 170±25 | ? |
| Amalthea | Jupiter | 135 x 83 x 75 | 1.8:1 |
| Phoebe | Saturn | 110 | Round to 9%, 7% relief |
| Janus | Saturn | 110 x 100 x 80 | 1.4:1 |

**Figure 1.** Which moons are spherical?
This mosaic displays 9 Solar System moons where all of the images have been degraded in resolution as needed such that they are roughly 8 pixels along the diameter (just as for the solitary resolved Nereid image). All nine mosaic panels have the moons around 96° solar phase angle, solar illumination from the left side, and a uniform grey scale stretched to go from the background to the peak pixel. None of the images have been processed (for example, with smoothing or selective windowing) as this can only add assumption-dependant properties. As such, these images are all directly comparable and are of equal efficiency in determining whether the moon is spherical or nonspherical. The reader is invited to examine these images and to decide which panels display nearly spherical moons and which panels display greatly-nonspherical moons. Only after this examination should the reader consult Table 6 to see how well they scored. If a perfect set of shape assignments is made, then we might have confidence that the *Voyager 2* image of Nereid can be used to decide whether Nereid is spherical. Alternatively, in the case as we find by polling colleagues, if the shape assignments are wrong about half the time, then we realize that the *Voyager 2* image places little useful constraint on the shape of Nereid.

**Figure 2.** Nereid Phase Curves from 1987-2006.
The nine panels of this figure display our observed phase functions for 1987-1990, 1993-1997, 1998, 1999, 2000, 2003, 2004, 2005, and 2006. Our best-fit Hapke model phase curve (from equation 1) is overplotted as the curved line. This plotted phase curve (from equation 1) is identical in all panels. The first two panels only display the nightly averages.

**Figure 3.** Combined phase curve for 'constant' years.
For the years 1998, 1999, 2000, and 2006, Nereid's observed phase function shows no significant variability. In these years, the phase curves are essentially identical. This figure shows the four years' data plotted together. The curve running down the middle of the points shows continuous curvature. The Hapke model of opposition surge brightness does provide a good fit to Nereid's phase function (the smooth curve). We can only get acceptable fits by having coherent backscattering dominating over shadow-hiding.

**Figure 4.** 1987-2006 light curve with all distance and geometry effects removed.
The quantity $m-5*Log_{10}[r\Delta/900]-m_{opp}$ is the residual in our observed magnitudes. The variations that remain are only those intrinsic to Nereid, for example due to rotational modulation and perhaps precession changes in the rotation. We plot this $m_{residual}$ for all data from 1987-2006. We can see that Nereid had large amplitude variations before 1991, was a quarter of a magnitude brighter in 2004, and has its amplitude of variation changing from year-to-year. This plot also shows the amount and distribution of our data on Nereid over the last twenty years.

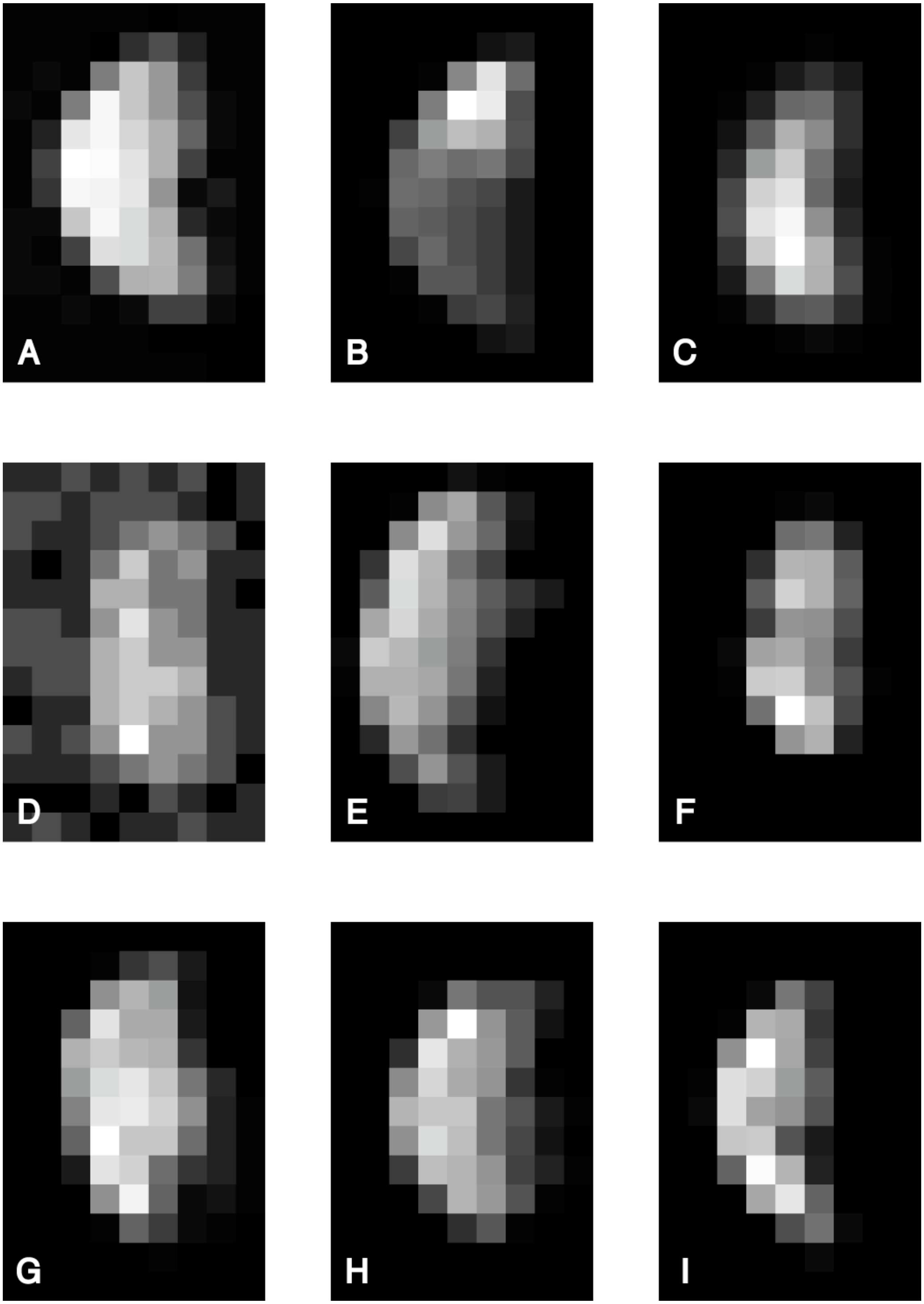

Figure 1

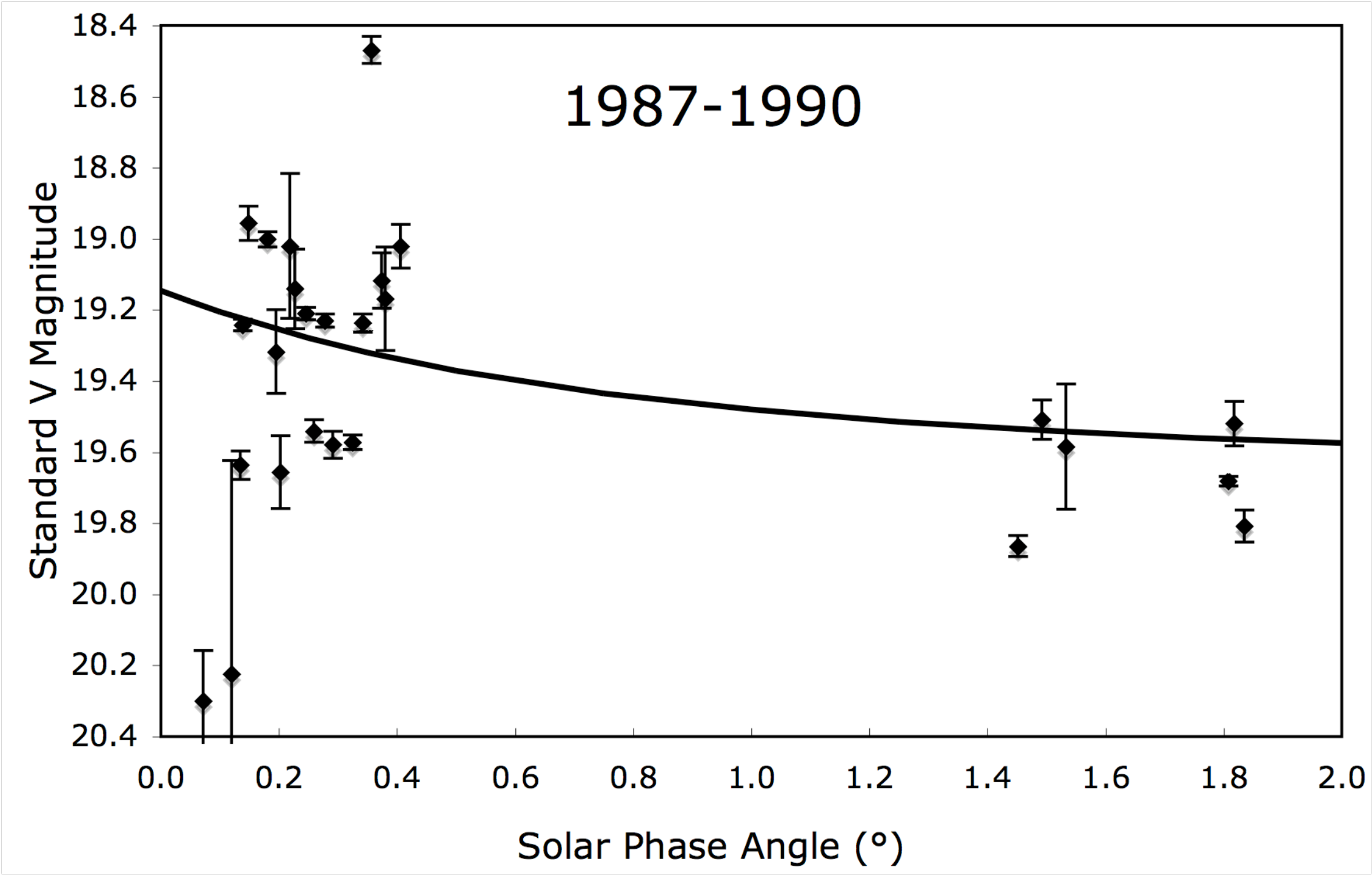

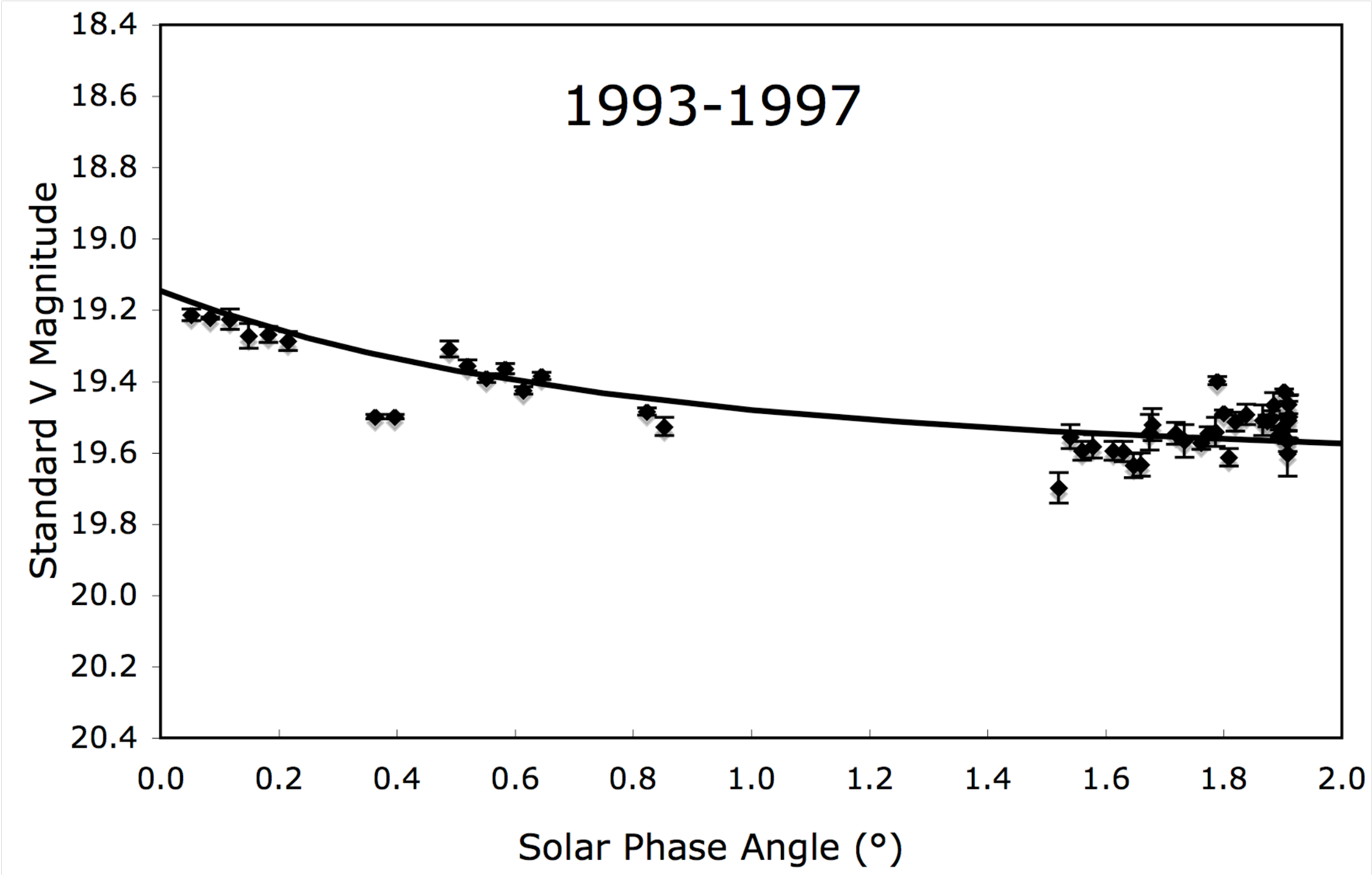

Figure 2 panels 1 and 2

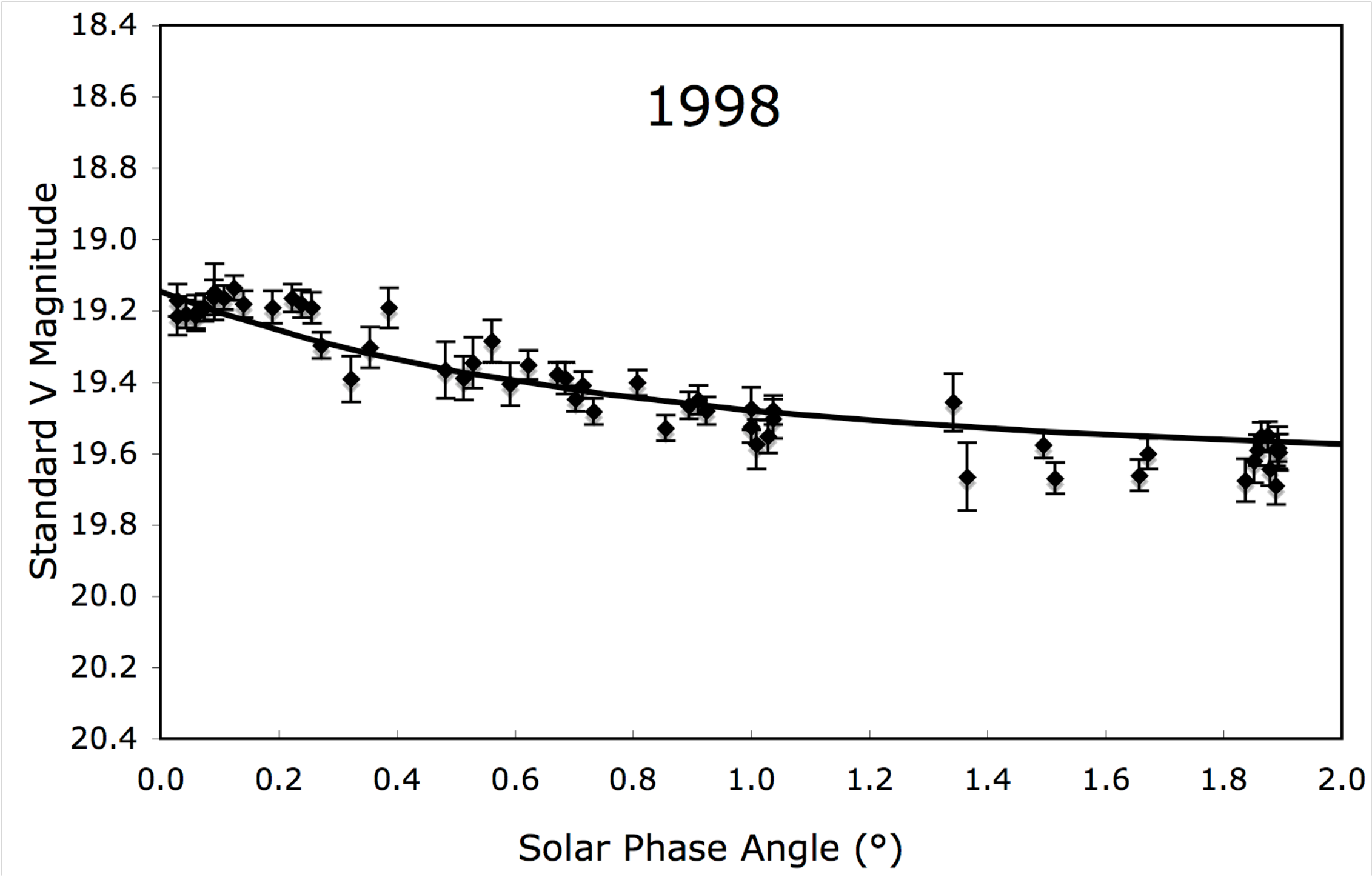

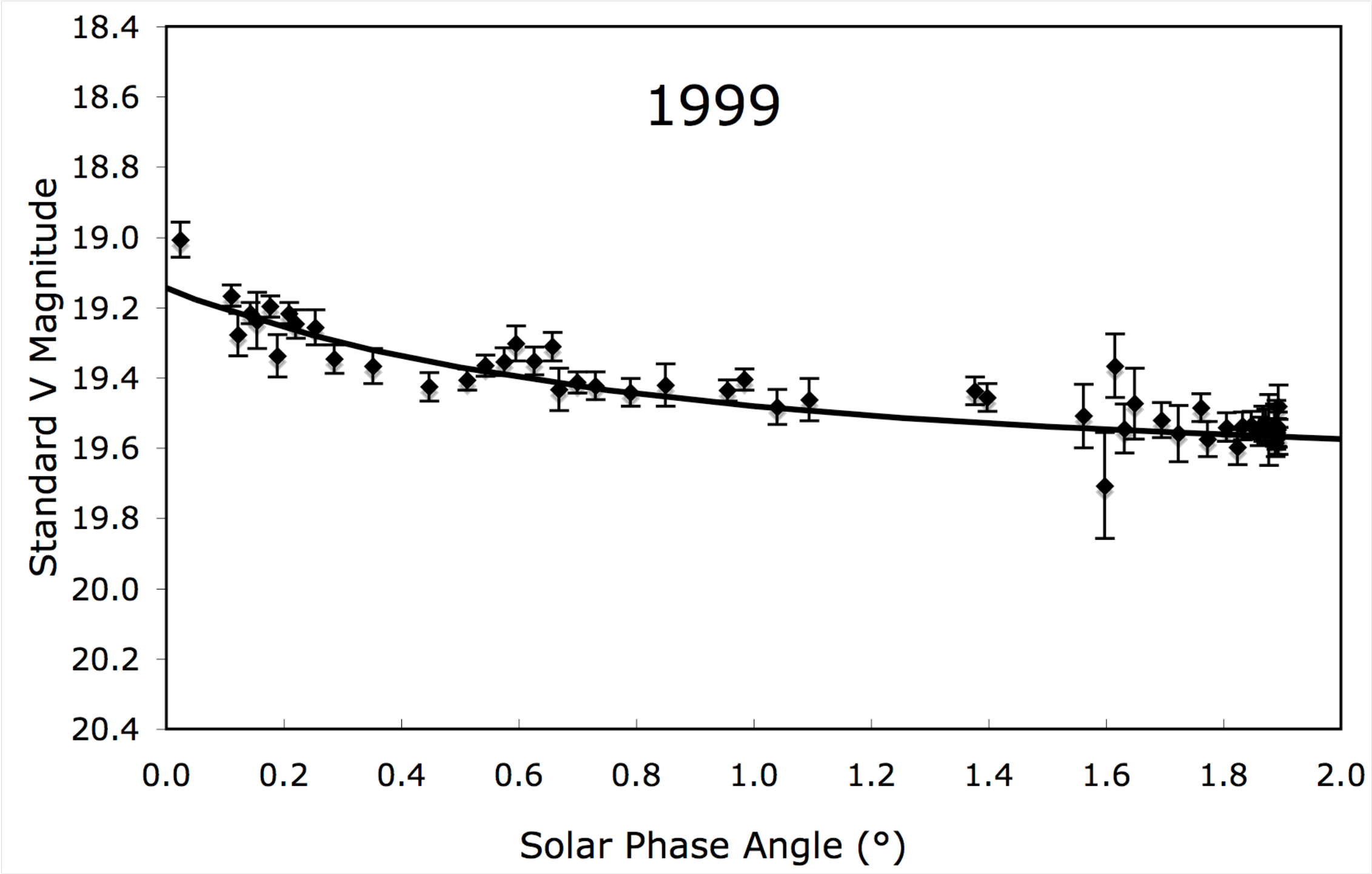

Figure 2 panels 3 and 4

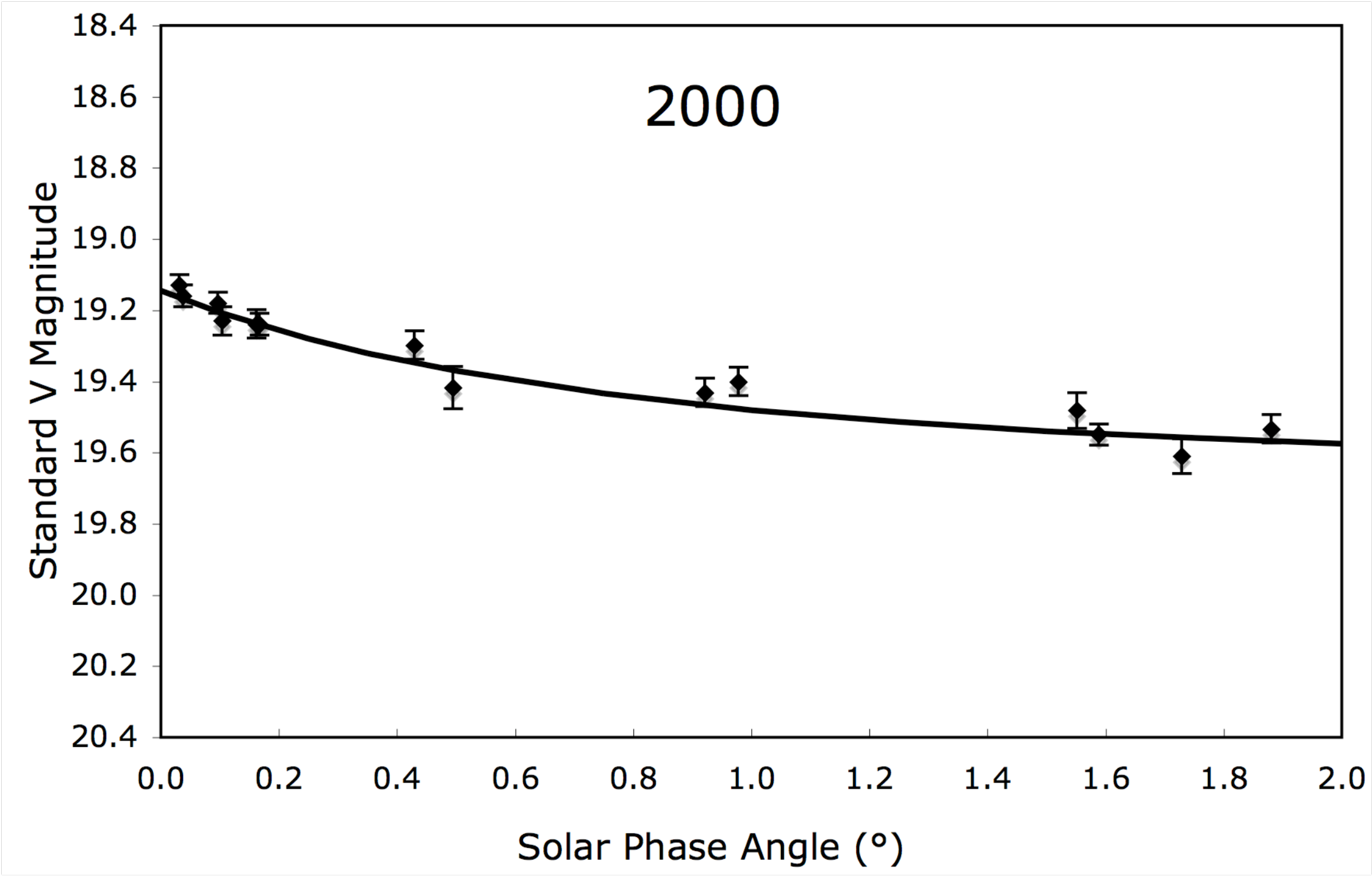

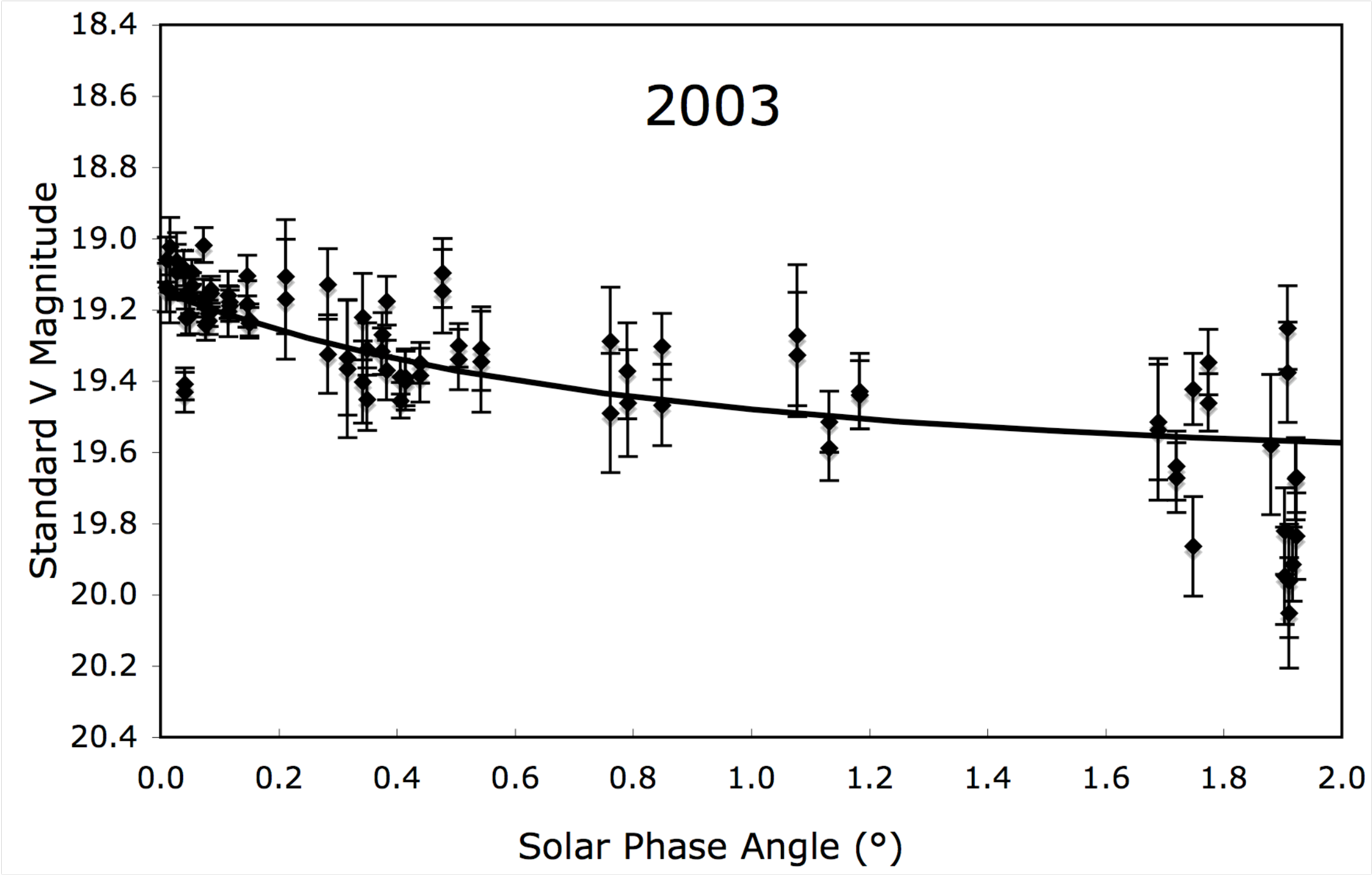

Figure 2 panels 5 and 6

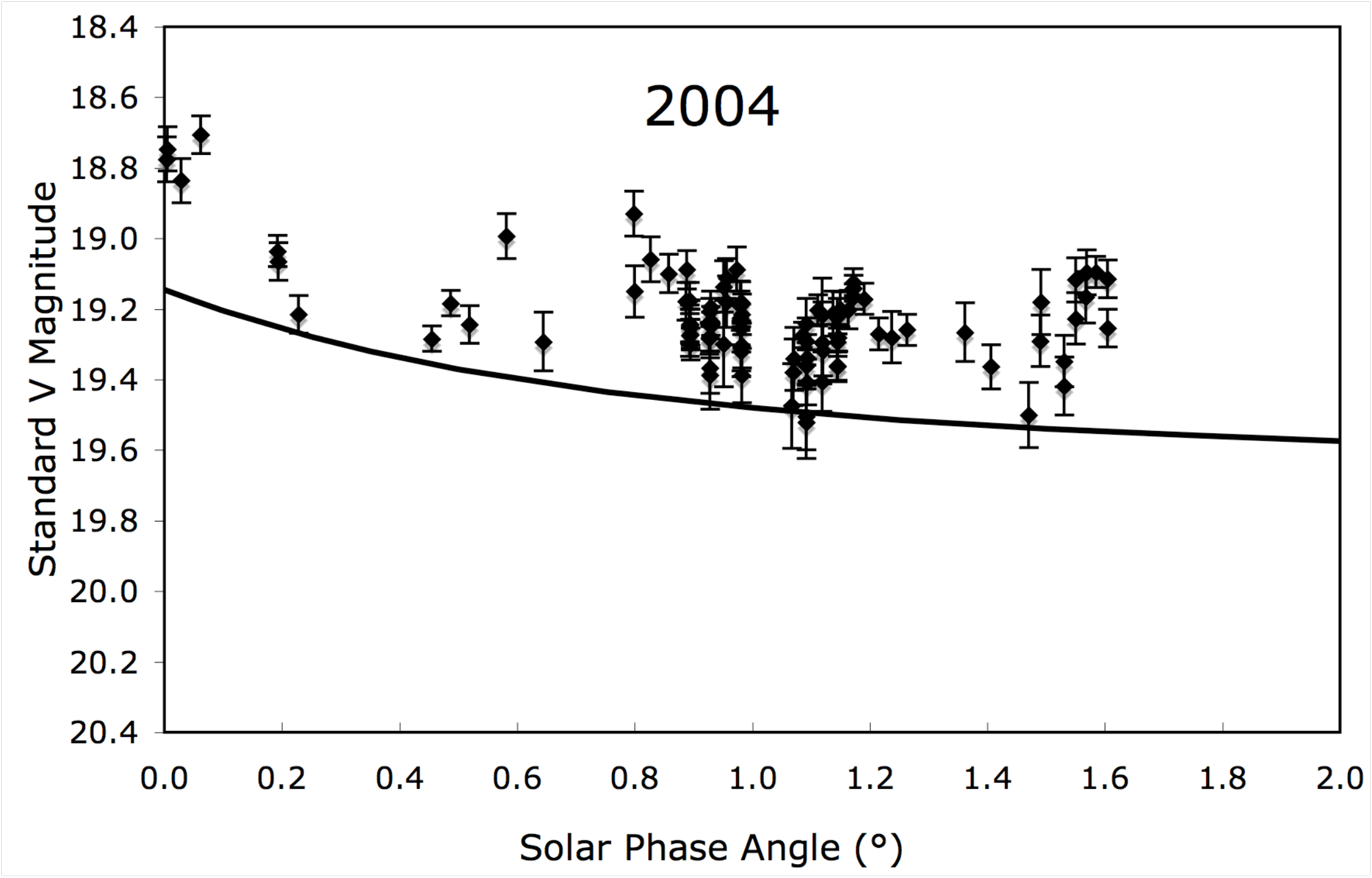

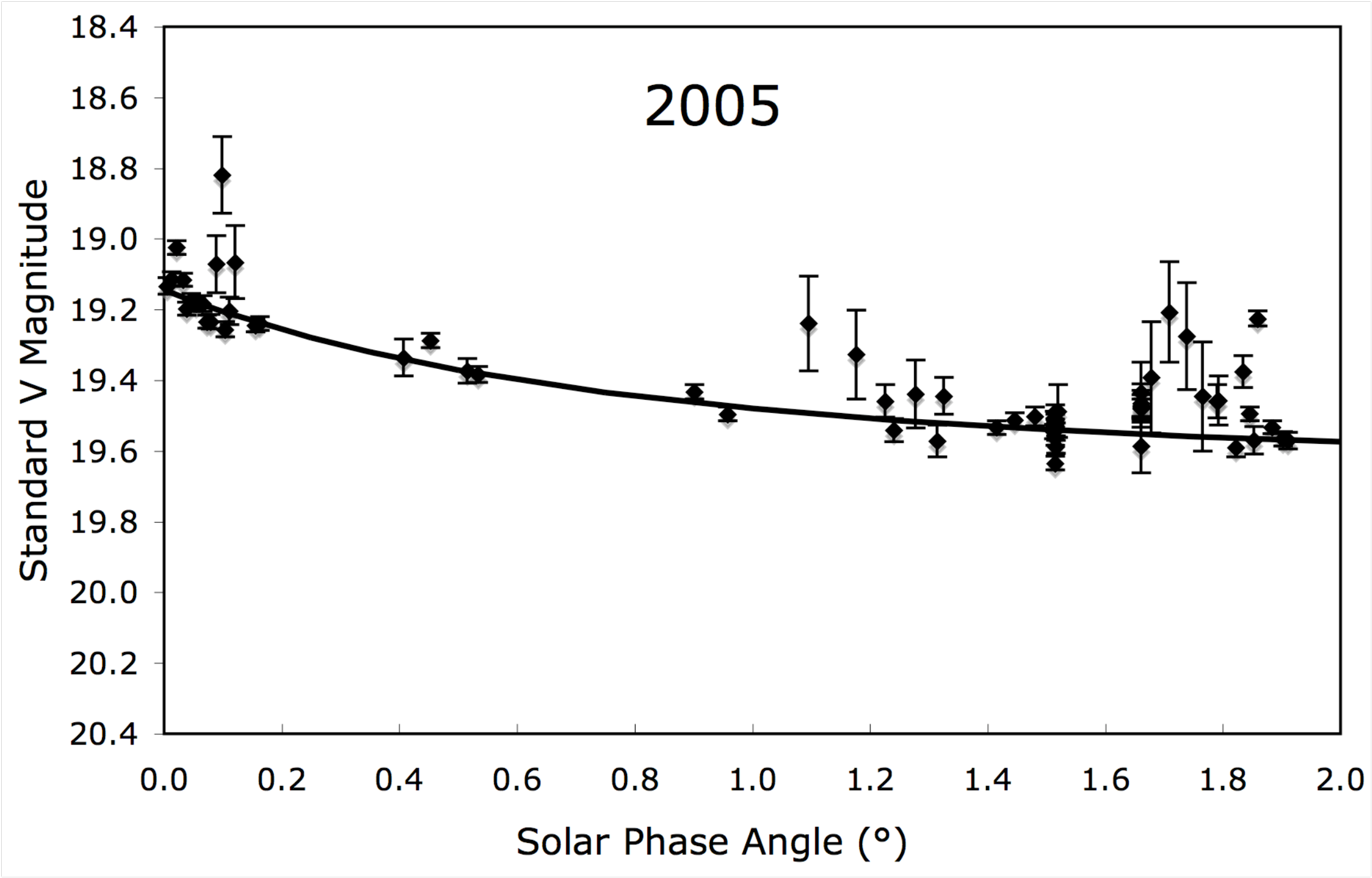

Figure 2 panels 7 and 8

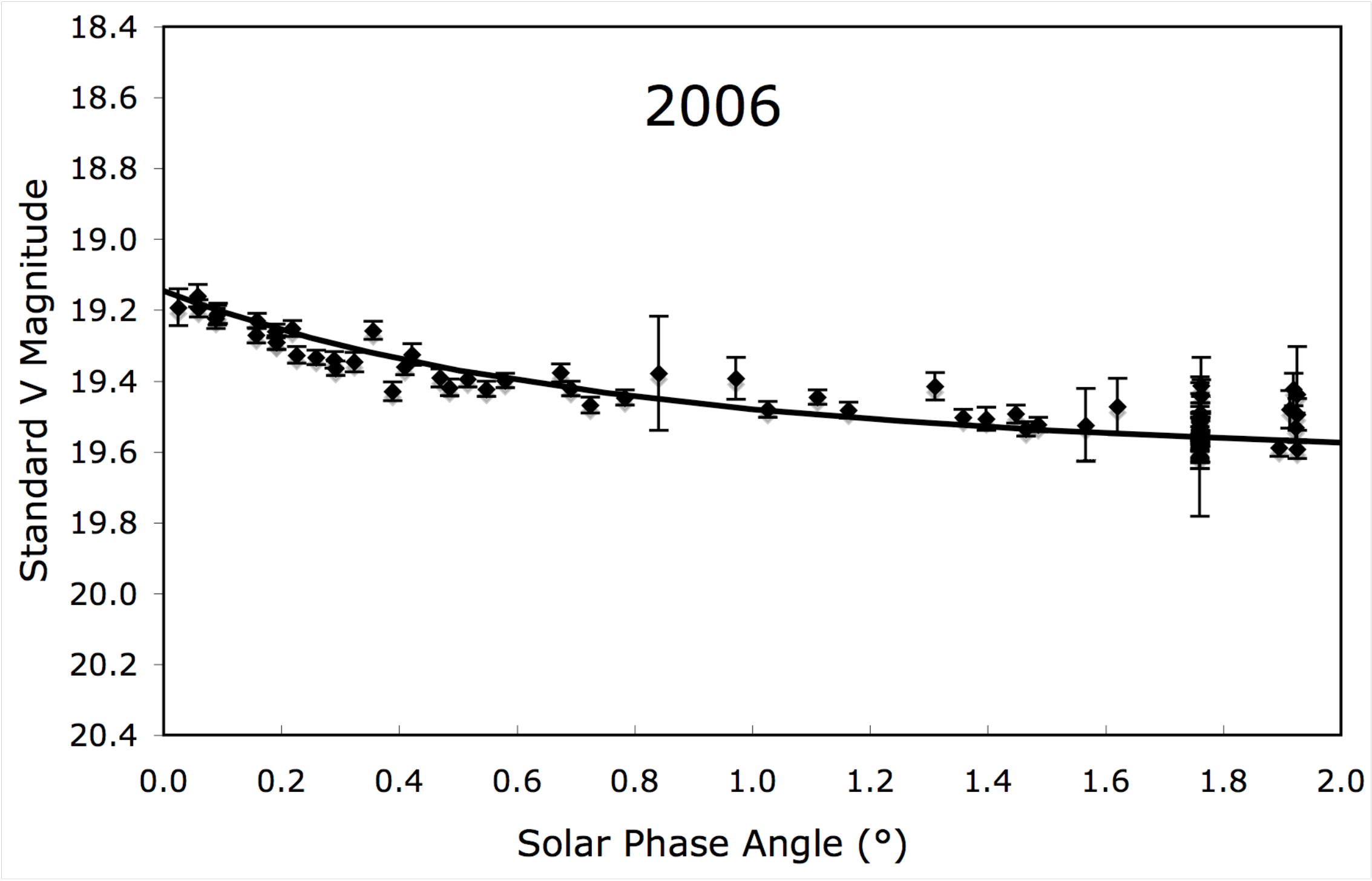

Figure 2 panel 9

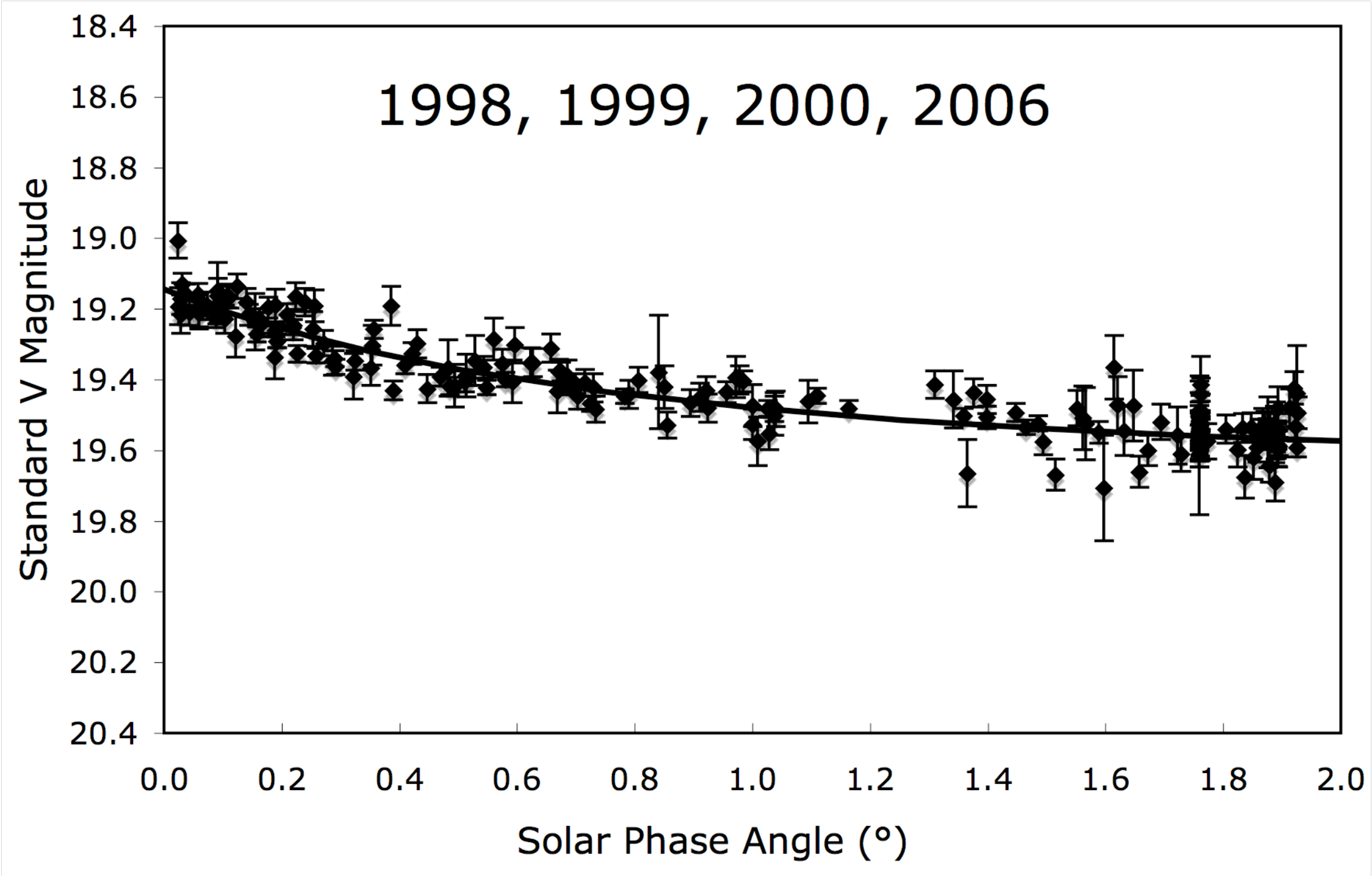

Figure 3

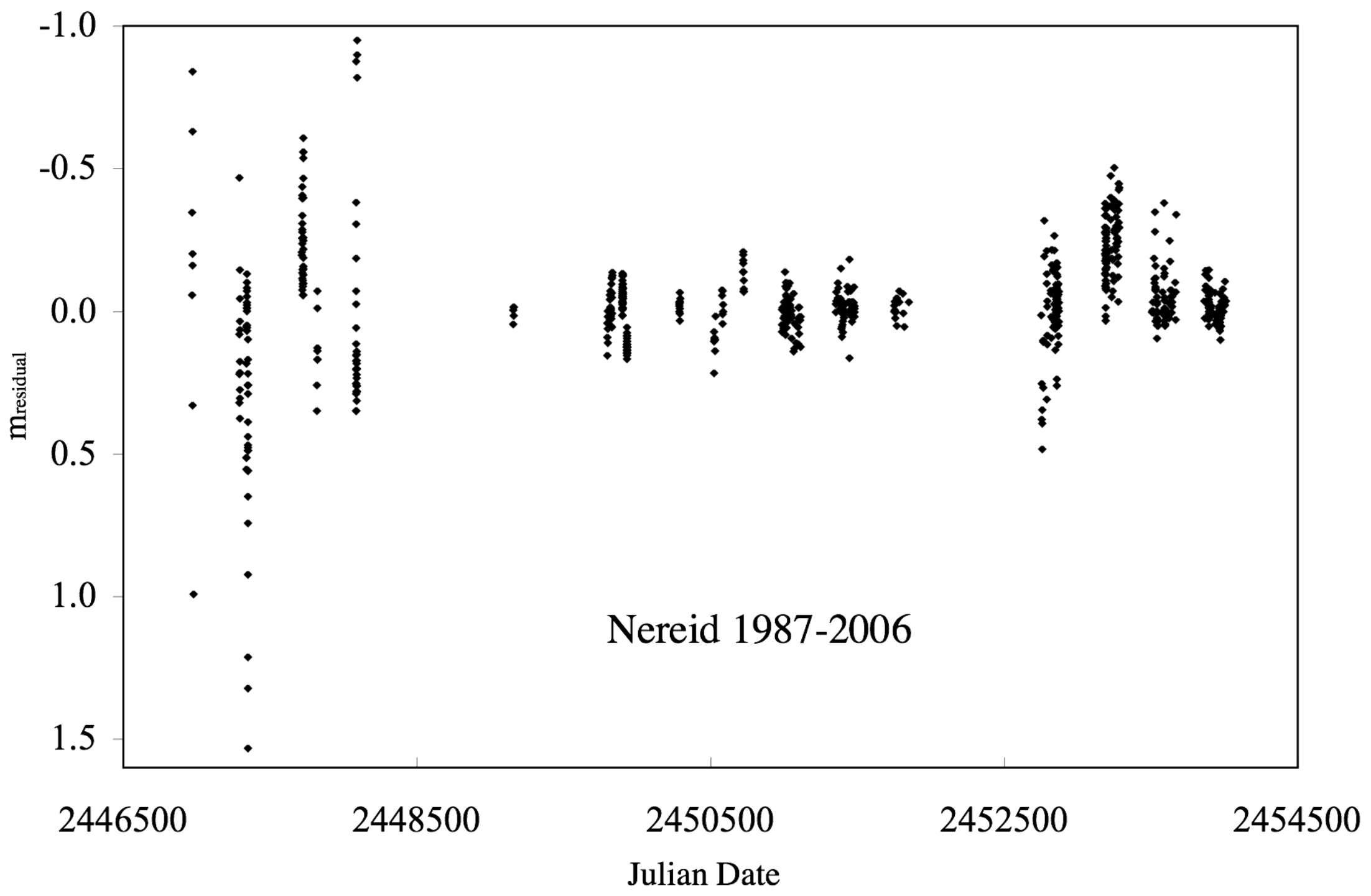

Figure 4